\pdfoutput=1
\documentclass[prd,superscriptaddress,preprint,eqsecnum,nofootinbib,amsmath,amssymb,preprintnumbers,tightenlines,longbibliography]{revtex4-1}
\usepackage[utf8]{inputenc} 
\usepackage{afterpage}
\usepackage{mathrsfs}
\usepackage{graphicx}
\usepackage{bm}
\usepackage{paralist}
\usepackage[colorlinks,linkcolor=blue,citecolor=blue]{hyperref}
\usepackage{mathtools}

\def\be{\begin{eqnarray}}
\def\ee{\end{eqnarray}}
\def\st{\begin{equation}}
\def\stp{\end{equation}}

\def\t{\tilde }

\def\p{{\bf p}}
\def\P{{\mathcal P}}

\def\bra{\langle}
\def\ket{\rangle}
\def\bbra{\langle\!\langle}
\def\kket{\rangle\!\rangle}

\def\ln{\mbox{ln}}

\def\tR{\tau_{\rm rel}}
\def\I{{\mathcal I}}
\def\P{{\mathcal P}}

\def\S{{\mathcal S}}

\newcommand{\rmd}{{\rm d}}

\def\Eq#1{Eq.~(\ref{#1})}
\def\Eqs#1{Eqs.~(\ref{#1})}
\def\App#1{Appendix~\ref{#1}}
\def\Fig#1{Fig.~\ref{#1}}
\def\Sect#1{Section~\ref{#1}}

\newcommand\bal{\begin{align}}
\newcommand\eal{\end{align}}

\begin{document}

\title{Onset of hydrodynamics for a  quark-gluon plasma from the evolution of moments of distribution functions}

\author{Jean-Paul Blaizot}
\email{jean-paul.blaizot@cea.fr}
\affiliation
{
	Institut de Physique Th{\'e}orique, Universit\'e Paris Saclay, 
        CEA, CNRS, 
	F-91191 Gif-sur-Yvette, France\\
}
\author{Li Yan} 
\email{liyan@physics.mcgill.ca}
\affiliation
{
	Institut de Physique Th{\'e}orique, Universit\'e Paris Saclay, 
        CEA, CNRS, 
	F-91191 Gif-sur-Yvette, France\\
}
\affiliation
{
Physics Department,
McGill University\\
3600 rue University
Montréal, QC
Canada H3A 2T8
}
\begin{abstract}

The pre-equilibrium evolution of a quark-gluon plasma produced in a heavy-ion
collision is studied in the framework of kinetic theory. 
We discuss the approach to local thermal equilibrium, and the onset of hydrodynamics,  in terms of a particular set of moments of the distribution function. 
These moments quantify the momentum anisotropies to a finer degree than the commonly used ratio of longitudinal to  transverse pressures. They are found to be in direct correspondence 
with  viscous corrections of  hydrodynamics, and provide therefore an alternative measure of these corrections in terms of the distortion of the momentum distribution. 
As an application, we study the evolution of these moments by solving the 
Boltzmann equation for a boost invariant expanding system, first analytically in the relaxation time approximation, and then 
 numerically for  a quark-gluon plasma with a collision kernel given by leading order $2\leftrightarrow 2$ QCD matrix elements in  the small angle approximation.

\end{abstract}

\maketitle

\section{Introduction}

The evolution of the quark-gluon plasma (QGP) produced in high energy heavy-ion collisions is well  described by the equations of 
relativistic hydrodynamics  including viscous corrections  (see~\cite{Heinz:2013th} for a recent review). The success of such hydrodynamic  descriptions 
suggests that  that the system of quarks and gluons which emerge shortly after the collisions is brought close to local equilibrium on a relatively short time scale, a process commonly referred to as thermalization. Understanding the detailed mechanisms by which such a thermalization occurs 
remains a 
theoretical challenge. At very early times, one may argue that the dynamics is dominated by classical color fields 
and the system evolution is governed by the classical Yang-Mills 
equations~\cite{Gelis:2013rba,Berges:2013fga}. At a time scale of order $1/Q_s$, where $Q_s$ is the saturation momentum, 
the system may be  sufficiently dilute for
kinetic theory to become applicable~\cite{Mueller:2002gd}. 
Kinetic theory  naturally fills the gap between the dynamics of classical field and that of dissipative fluids, and it offers the possibility to follow in details how the  pre-equilibrium system evolves into a state of quasi local equilibrium well accounted for by viscous hydrodynamics. This is the framework that we shall consider in this work (see e.g. \cite{Kurkela:2011ti,Kurkela:2014tea,Blaizot:2013lga,Blaizot:2014jna,Xu:2014ega} for some recent representative works, and more specifically \cite{Kurkela:2015qoa,Keegan:2015avk} for the   analysis 
of the onset of hydrodynamics
in a weakly coupled system using kinietic theory).

The transition from kinetic theory to hydrodynamics is commonly achieved by taking suitable moments of the kinetic equations, with lower moments encompassing conservation laws and higher moments various dissipative effects. For some particular geometries, it is possible to recast the solution of the Boltzmann equation in terms of an infinite hierarchy of equations for a particular set of moments of the 
distribution function (see e.g.~\cite{Bazow:2016oky}). Provided this infinite hierarchy can be limited to the first few moments, this technique could represent a convenient alternative to the direct solution of the kinetic equation. 
Aside from this aspect, there is another interest in using moments. 
Defined in terms of integrals of the phase-space 
distribution function with suitable weights, the moments allow us to focus on the relevant (typically long wavelength) information, and wash out the irrelevant  (short wavelength) one 
from the distribution function, thereby automatically implementing a strategy akin to that of effective field theories.

 In this paper, we introduce a specific set of moments, defined as weighted integrals of the momentum distribution function $f(p)$, ${\cal L}_n\propto \int_p p^2 P_{2n}(\cos\theta) f(p)$, where $P_{2n}$ is a Legendre polynomial, and $\cos\theta=p_z/p$ with $p_z$ the longitudinal momentum of a particle with total momentum $p$.   These moments capture the deviation of the momentum distribution from an isotropic distribution and are therefore damped as the system approaches local equilibrium. They are tailored to take into account the (strong) effect of the longitudinal expansion.
 In addition, these moments are found to be in 
correspondence with viscous corrections in hydrodynamics, order
by order in a gradient expansion. They can therefore provide an alternative view of the viscous corrections, in terms of the damping of high multipoles of the momentum distribution as the system approaches local equilibrium. Their relative magnitudes can be used as an indicator of the onset of hydrodynamics. 
In this work, we shall provide a description of the evolution of these moments, obtained by solving the Boltzmann equation for a 
longitudinally expanding system. The usefulness of these moments as a practical tool in solving the kinetic equation will be addressed in a separate publication.
 
This paper is organized as follows. The moments ${\cal L}_n$, and their usefulness in a boost invariant setting, are introduced in 
\Sect{sec:sec2}. In this section, we 
 also present some of the major results obtained in this work, concerning the 
correspondence between the ${\cal L}_n$'s and  the transport coefficients used in second order  viscous conformal hydrodynamics \cite{Baier:2007ix}. Relations to higher order viscous corrections are also derived. Then, in \Sect{sec:sec3}, we discuss  the evolution of the moments in the pre-equilibrium
stage towards local equilibrium using kinetic theory.  To do so we solve the Boltzmann equation in a boost invariant setting, using two different approximations for the collision kernel. In the first case, we use 
the relaxation time approximation, with a constant relaxation time, and obtain analytical solutions that allow us to verify the general relations to 
viscous hydrodynamics established in \Sect{sec:sec2}.
Then,  in \Sect{sec:sec3b}, we consider a more realistic setting: we use leading order $2\leftrightarrow 2$ QCD matrix elements for the collision kernel, and solve the corresponding Boltzmann equation in the small scattering angle approximation in order to study the evolution of the moments. Summary and conclusions are given in \Sect{sec:sec4}.

\section{Formulation of moments in  expanding systems}
\label{sec:sec2}

In this section, after a  brief review of the main features of expanding boost invariant systems, we define a set of
 moments of the distribution function that are suited to the description of such systems. We show that, near  the hydrodynamic regime, these
moments  are in correspondence to the viscous corrections that emerge from a gradient expansion.

\subsection{Kinetic theory in a boost invariant expanding system}
The quark-gluon plasma produced in the very early stages of a heavy ion collision  experiences a strong expansion along the collision axis (referred to as the longitudinal direction). A simple description of the system (Bjorken flow) is obtained when one assumes boost invariance in the longitudinal direction and translational invariance in the transverse directions \cite{Bjorken:1982qr}. It becomes then convenient to use in place of the usual  space-time coordinates, the proper time 
$\tau=\sqrt{t^2-z^2}$ and the space-time rapidity $\tan^{-1}(z/t)$ , where $z$ is the longitudinal coordinate and $t$ the time. 
In fact,  boost invariance makes it possible to focus on a slice  of the fluid located around the plane $z=0$, where $\tau=t$.   There, the distribution function depends only on time and the three components of the momentum, $f(t,p_T,p_z)$, and it obeys the  kinetic equation
\be\label{eq:kinetic}
\frac{\partial f}{\partial t}-\frac{p_z}{t}\frac{\partial f}{\partial p_z}=-{\cal C}[f],
\ee
with ${\cal C} $ the collision term\footnote{
For simplicity, 
we consider in this section a single-species system. In Sect.~\ref{sec:sec3b} we shall introduce distinct distributions for quarks and gluons and take proper care of degeneracy factors}.

Averages of various physical quantities with  the phase space distribution function play an important role in this paper, and we shall denote them with double brackets\footnote{Throughout this paper, 
we use
bold lower case letters to denote a three vector,  such as $\p$, with the magnitude of 
the vector denoted by the corresponding normal lower case letter,
e.g., $|\p|=p$.}
\be
\bbra\ldots\kket \equiv
\int_p \ldots f\, ,\qquad 
\int_p\equiv
\int \frac{d^3 \p}{(2\pi)^3 p^0},
\ee
For instance, the energy density is given by 
$
e(t)= \bbra (p^0)^2 \kket =\bbra p^2 \kket, 
$
where we have used $p_0=p$, assuming massless particles. 

By multiplying Eq.~(\ref{eq:kinetic}) by $p^2$, and integrating over momentum, we obtain,
\be\label{eq:energy}
\frac{\rmd e}{\rmd t}+\frac{e(t)+{\cal P}_L(t)}{t}=0,
\ee
where we have used the fact that the collisions conserve energy, so that the contribution of the collision term vanishes. The quantity ${\cal P}_L$ is the longitudinal pressure, 
\be
{\cal P}_L(t)\equiv \bbra p_z^2 \kket=-\frac{\rmd ( t e(t))}{\rmd t}.
\ee
We define similarly the transverse pressure
\be
\P_T\equiv\frac{1}{2}\bbra p_T^2\kket\equiv\frac{1}{2}\int_p (p_x^2+p_y^2) f(\p)=\frac{1}{2t}\frac{\rmd(t^2 e(t))}{\rmd t},
\ee
where in the last equality, we have used $e(t)={\cal P}_L+2{\cal P}_T$ and Eq.~(\ref{eq:energy}). 

In the hydrodynamic regime, i.e. when local equilibrium is achieved,  the pressure is isotropic and simply related to the energy density,  ${\cal P}={\cal P}_L={\cal P}_T=e/3$, and Eq.~(\ref{eq:energy}) becomes a closed equation for the energy density. It yields $e(t)\sim 1/t^{4/3}\sim T^4$, with $T(t)$ the local temperature. Before reaching this regime, viscous corrections need to be taken into account. The first order correction involves the shear viscosity $\eta$, and yields a relaxation equation for the difference of the pressures
\be\label{eq:hydro_mom1}
{\cal P}_T-{\cal P}_L=2\frac{\eta}{t}.
\ee
Viscous corrections are accompanied by an entropy increase, given in leading order by the equation (with $Ts=e+{\cal P}$, where $s$ is the entropy density and $T$ the temperature)
\be
\frac{\rmd (ts)}{\rmd t}=\frac{4\eta}{3 tT}.
\ee
Note that $ts$ represents the total entropy in a (expanding) covolume, or equivalently the entropy density in the transverse plane. In the absence of viscosity, this is contant. 

Our purpose in this paper is to explore specific features of the approach to the hydrodynamical regime, studying  in particular how the deviations of the momentum distribution from an isotropic distribution are damped.


\begin{figure}
\begin{center}
\includegraphics[width=0.56\textwidth] {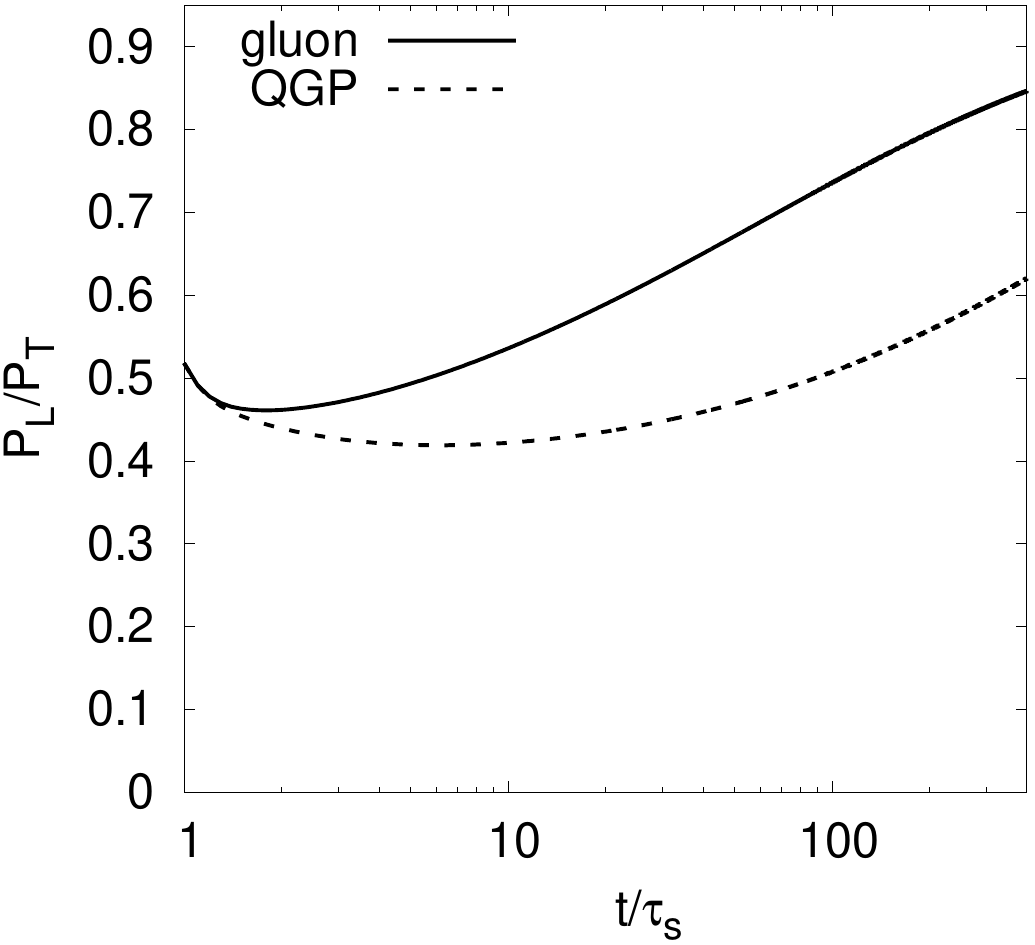}
\caption{ (Color online)
Time evolution of the pressure ratio
$\P_L/\P_T$ 
obtained from the numerical solution of the Boltzmann equation in the small angle approximation, 
for a  pure gluon system (solid line) and a quark-gluon plasma (QGP, dotted line).  Here $\tau_s$ is a natural microscopic time scale, proportional  to the inverse of the saturation momentum $Q_s$ (see \Sect{sec:sec3b} for details).
\label{fig:iso}
}
\end{center}
\end{figure}

\subsection{Thermalization, isotropization and moments}

Quite generally, the effect of collisions is to wash out the anisotropy of the 
momentum distribution, leading eventually to  a spherically symmetric distribution. 
In the case of expanding boost invariant systems, this effect is counterbalanced 
by the strong longitudinal expansion. The competition between the two effects
is commonly investigated in terms of the ratio 
between the longitudinal pressure
$\P_L$ and the transverse pressure $\P_T$, local equilibrium being established when
$\P_L/\P_T=1$. As an illustration, and anticipating on the discussion of 
\Sect{sec:sec3b}, we show in \Fig{fig:iso} the 
typical evolution of  $\P_L/\P_T$ obtained from the 
numerical solution of the Boltzmann equation, with leading order $2\leftrightarrow 2$ scatterings of QCD matrix elements, 
in the small scattering angle approximation.
The trend towards isotropization is clearly visible. However, this is a slow process, 
and  complete local thermal equilibrium is not reached in the time span of the simulation, 
with quark production delaying the approach to equilibrium even further as 
compared to the case of a purely gluonic plasma.\footnote{Some readers may find it surprising to see this trend towards isotropization arise from elastic scattering alone, as it seems to go against a prediction from the bottom-up scenario \cite{Baier:2000sb} emphasizing the role of inelastic collisions in isotropization. The parametric analysis of the bottom-up scenario  relies however on values of the coupling constant much smaller than those used in the  present simulation ($\alpha_s\approx 0.3$). In fact, Tanji and Venugopalan \cite{Tanji:2017suk} have recently completed a systematic study of the same equations as ours, in which they show explicitly how deviations form the bottom-up scenario develop as the coupling increases.    } Such a slow approach to isotropy of the 
momentum distribution seem to be quite generic for a longitudinally expanding 
system undergoing Bjorken flow~\cite{Keegan:2015avk}.  
However, it has also been realized that the complete isotropy of the pressures,
characterized by $\P_L/\P_T=1$, may not 
be necessary for hydrodynamics to be applicable, since viscous corrections can accommodate rather large differences between ${\cal P}_L$ and ${\cal P}_T$. 
This was first revealed by strong coupling 
calculations \cite{Heller:2011ju}, where it was shown that viscous hydrodynamics can 
handle  pressure ratios $\P_L/\P_T\gtrsim 0.5$. 
This has been later verified in a number of calculations (see Ref.~\cite{Keegan:2015avk} and 
references therein). 

In this paper, in order to describe more details of the isotropization of an 
expanding quark-gluon plasma,  we introduce the following moments of the distribution function
\be
\label{eq:def_moments}
{\cal L}_n\equiv \bbra p^2P_{2n}(\cos\theta)\kket\equiv \int_p p^2P_{2n}(\cos\theta) f(\p),
\ee
where $P_{2n}$ is a Legendre polynomial of order $2n$,  and $\cos\theta = p_z/p$. 
For an expanding system
with Bjorken geometry, odd order moments vanish as a consequence
of the invariance of the distribution function under parity.
There are two distinct advantages of these moments. First,
except for the $n=0$ moment which 
corresponds to the energy density, as already mentioned, all higher order 
moments defined in \Eq{eq:def_moments}
naturally
quantify the details of the longitudinal momentum anisotropy. 
For instance,  the information concerning  $\P_L/\P_T$ 
is contained in the $n=1$ moment, 
\be\label{moment1}
{\cal L}_1=\bbra p^2P_{2}(\cos\theta)\kket=\P_L-\P_T\,.
\ee
More precisely, ${\cal L}_1\rightarrow0$ is equivalent to
$\P_L/\P_T\rightarrow1$. Similarly, the moments ${\cal L}_n$ of higher order are associated to finer 
structures of the momentum anisotropy of the distribution function. 
Second, the moments defined in \Eq{eq:def_moments}, with the specified weight $p^2$, 
are closely related to hydrodynamics through
Landau's matching condition 
\be
T^{\mu\nu}=\bbra p^\mu p^\nu\kket,
\ee
of which the moment ${\cal L}_1$ provides the simplest illustration (see Eq.~(\ref{eq:hydro_mom1})). 
These
  moments 
are therefore expected to acquire a physically transparent meaning  
at late times when the system approaches the hydrodynamic regime. Note that one may generalize the definition in \Eq{eq:def_moments} 
to  space-time symmetries other  than the boost invariant setup considered in this paper. For a system with spatial SO(3)
rotational symmetry, for instance, 
one could replace the Legendre polynomials with associated
Legendre polynomials to further characterize momentum 
anisotropies in the azimuthal direction.

\subsection{The moments ${\cal L}_n$ in the hydrodynamic regime}

We now analyze more precisely the correspondence of the moments (\ref{eq:def_moments}) 
to viscous hydrodynamics. At this point covariant notation is  helpful. Four vectors are denoted by  normal upper case letters. Thus, for instance, $U_\mu$ is the fluid  four-velocity, normalized so that $U^2=-1$. The  operator  $\Delta_{\mu\nu}
=g_{\mu\nu}+U_\mu U_\nu$, with $g_{\mu\nu}$ the metric tensor\footnote{We use the Minkowski metric signature $(-,+,+,+)$. More generally we refer to \cite{Teaney:2013gca,Yan:2012jb} for more details on the notation.}, projects on the space orthogonal to $U^\mu$. The energy momentum tensor is written as 
\begin{align}
\label{eq:brsss}
T^{\mu\nu} = (e+\P)U^\mu U^\nu + \P g^{\mu\nu} + \pi^{\mu\nu}\,
\end{align}
where $ \pi^{\mu\nu}$ represents the viscous corrections. In leading order, $ \pi^{\mu\nu}=-\eta\sigma^{\mu\nu}$, where $\eta$ is the shear viscosity, and  $\sigma^{\mu\nu}=2\bra \nabla^\mu U^\nu \ket\,,
$
with $\nabla^\mu=\Delta^{\mu\nu}\partial_\nu$. Here, and in the following, single brackets $\bra \ldots\ket$ around  a 
tensor  implies that the  tensor has been made symmetric,
traceless and transverse to $U^{\mu}$. For simplicity, we assume throughout this
work that the fluid is made of massless constituents, so that conformal symmetry implies that the energy-momentum tensor is traceless, $T^{\mu}_{\;\;\mu}=0$. 

Hydrodynamics effectively applies to systems whose evolution is dominated by long wavelength modes. 
Corrections to ideal hydrodynamics are then naturally searched for in a gradient expansion. The successive terms in such an expansion can be obtained by applying the Chapman-Enskog technique.  

As an illustration of the method, let us repeat the derivation of \Eq{eq:hydro_mom1} within the relaxation
time approximation for the collision term. 
Assuming that the viscous correction corresponds to a small deviation $\delta f$ of the  phase-space distribution function from its local equilibrium value $f_{\rm eq}$, 
we linearize the Boltzmann equation and obtain, in covariant notation,
\be
{P^\mu} \partial_\mu (f_{\rm eq} + \delta f)
={P\cdot U}\frac{\delta f}{\tR}\,.
\ee
where $\tR$ is the relaxation time. 
Note that 
$P\cdot U$ reduces to $-P^0$ in the local rest frame of a fluid
cell.
We shall allow the relaxation time to depend on the energy of the particle, i.e, we assume $\tR=\tR(P\cdot U/T)$, where $T$ is the local temperature (which enters the local equilibrium distribution). Various ansatz have been considered in the literature, in particular a so-called ``linear'' ansatz, corresponding to a constant relaxation time, and a ``quadratic'' ansatz, corresponding to a linear dependence of $\tR $ on the energy \cite{Dusling:2009df}. The origin of this terminology is that, in the first case, $\delta f/f_{\rm eq}\sim p/T$, while $\delta f/f_{\rm eq}\sim (p/T)^2$ when $\tR$ is linear in $p$.
Note that the present  analysis does not rely on the specific dependence of the  relaxation time upon energy. 

To proceed, we find it convenient to  define
\be
\t C_p = -T^2\frac{\tau_{\rm rel}(P\cdot U/T)}{P\cdot U}\,,
\ee
so that 
\be
\label{eq:tR_hydro}
P^\mu \partial_\mu (f_{\rm eq}+\delta f) = -\frac{T^2}{\t C_p} \delta f\,.
\ee
Since the local equilibrium distribution is a function of  $P\cdot U/T$,
 the  effect of the operator $P^\mu \partial_\mu$ when acting on $f_{\rm eq}$ is to
generate the structure $P^\mu P^\nu \sigma_{\mu\nu}$, i.e.,
\be
P^\mu \partial_\mu f_{\rm eq}(P\cdot U/T) = -f_{\rm eq}'\frac{P^\mu P^\nu \sigma_{\mu\nu}}{2T}
+O(\nabla^2)\,,
\ee
where the prime on $f_{\rm eq}$ indicates a derivative with respect to $P\cdot U/T$. 
Ignoring the derivative of $\delta f$ in the left hand side of Eq.~(\ref{eq:tR_hydro}), one then identifies the first order viscous correction to the phase-space distribution function 
\be
\label{eq:df1}
\delta f=\t C_p f_{\rm eq}' \frac{P^\mu P^\nu \sigma_{\mu\nu}}{2T^3}
+ O(\nabla^2),
\ee
as well as the first viscous correction to the energy momentum tensor, 
\be\label{viscouspimunu}
\pi^{\mu\nu}= -\eta\sigma^{\mu\nu}=\int_p P^\mu P^\nu \delta f.
\ee
A simple calculation then allows us to determine the shear 
viscosity~\cite{Teaney:2013gca} 
\be
\label{eq:shear_viscosity}
\eta=-\frac{1}{15T^3}\int_{ p} \; p^4 \t C_p f_{\rm eq}'\,.
\ee
In the case of Bjorken flow, the contraction of the irreducible tensors in \Eq{eq:df1}  is easily calculated
in terms of the Legendre polynomials. One gets
\be
\label{eq:pps1}
P^\alpha P^\beta \sigma_{\alpha\beta}=\frac{4}{3t}
\left[p_z^2-\frac{1}{2}p_T^2\right]
=\frac{4}{3t}p^2P_2(\cos \theta).
\ee
By using this expression in  \Eq{eq:df1}, and the expression (\ref{eq:shear_viscosity}) of the shear viscosity, one can then calculate the $n=1$ moment, Eq.~(\ref{moment1}), and obtain  \Eq{eq:hydro_mom1}.

The higher order corrections are obtained iteratively along the same line. 
In the second order correction $\delta f^{(2)}$ needed to 
  the calculation of the transport coefficients of conformal viscous hydrodynamics \cite{Baier:2007ix}, the following 
new tensor structures appear~\cite{Teaney:2013gca,Yan:2012jb},
\begin{align}
\label{eq:tensors}
&P^\mu P^\nu P^\alpha P^\beta \!\,\bra \sigma_{\mu\nu}\sigma_{\alpha\beta}\ket\,,
\quad
P^\mu P^\nu P^\alpha \!\,\bra\sigma_{\mu\nu}\nabla_{\alpha}\ket\ln T\,,
\quad
P^\alpha P^\mu P^\nu \!\,\bra\nabla_\alpha \sigma_{\mu\nu}\ket\,,
\quad
P^\mu P^\nu \!\,\bra\sigma_{\mu\,}^{\;\,\lambda}\sigma_{\nu\lambda}\ket\,,\nonumber\\
&
P^\mu P^\nu \!\,\bra\sigma_{\mu\,}^{\;\,\lambda}\Omega_{\nu\lambda}\ket\,,\quad
P^\mu P^\nu \!\,\bra U^\mu\partial_\mu \sigma_{\mu\nu}\ket\,.
\end{align}
For Bjorken flow, the vorticity tensor $\Omega^{\mu\nu}=\frac{1}{2}\left[\nabla^\mu U^\nu -\nabla^\nu U^\mu \right]$ vanishes.
One can also prove that contractions of irreducible tensors of odd ranks do not contribute  
due to parity. 
Finally, the relevant structures 
in \Eq{eq:tensors} are those arising from the contraction of irreducible tensors of even ranks,
\begin{subequations}
\begin{align}
\label{eq:contractions}
P^\mu P^\nu P^\alpha P^\beta \!\,\bra\sigma_{\mu\nu}\sigma_{\alpha\beta}\ket
=&\frac{32}{35}\frac{p^4}{t^2}P_4(\cos\theta)\,,\\
P^\mu P^\nu \!\,\bra\sigma_{\mu\,}^{\;\,\lambda}\sigma_{\nu\lambda}\ket
=&\frac{8}{9}\frac{p^2}{t^2}P_2(\cos\theta)\,,\\
P^\mu P^\nu \!\,\bra D \sigma_{\mu\nu}\ket
=&-\frac{4}{3}\frac{p^2}{t^2}P_2(\cos\theta)\,.
\end{align}
\end{subequations}
One can thus rewrite the viscous corrections 
to the phase space distribution up to 
second order in the gradient expansion in terms of Legendre polynomials, 
\begin{align}
\label{eq:df_exp}
\delta f=\bigg[-\t \chi_p \t p^2 \left(\frac{2}{3t T}\right)
&+\t\chi_p'\t C_p\t p^4\left(\frac{8}{63t^2T^2}\right)
-\t\chi_p\t C_p\t p^3\left(\frac{8}{9t^2T^2}\right)+\ldots\bigg]P_2(\cos \theta)\nonumber\\
&+\bigg[\t\chi_p'\t C_p\t p^4 \left(\frac{8}{35t^2T^2}\right)+\ldots\bigg]P_4(\cos\theta)
+ \ldots\,,
\end{align}
where, for convenience, we have defined the dimensionless variables 
$\t \chi_p=f_{eq}' \t C_p$ and $\t p=p/T$.
Ellipses in the brackets of \Eq{eq:df_exp} stand for terms of higher 
power in $1/t T$. 

One recognizes in Eq.~(\ref{eq:df_exp}) a gradient expansion,  with $\delta f$
 expressed 
in powers of $1/t T$. The latter quantity may be viewed as a measure of the  Knudsen number, that is, as  the  ratio between a microscopic  and a macroscopic length scale characterizing the  fluid. Here the typical microscopic length scale is the inverse of the temperature, while the macroscopic length scale can be taken as the inverse of the local expansion rate, which, for  a medium system expanding according to Bjorken flow, is simply the time 
 $t$.  The expansion (\ref{eq:df_exp}) is also an expansion in Legendre Polynomials, the term multiplying $P_{2n}(\cos\theta)$ having a gradient expansion 
 starting with a leading contribution of order $n$, that is,  $\sim 1/(t T)^n$.\footnote{Note that the expansion (\ref{eq:df_exp}) contains also  terms proportional to $P_0(\cos\theta)$. These terms are not shown explicitly 
since they are not related to momentum anisotropies, which is our primary concern.}

More explicitly, the evolution of the moments of order $n=1$ and $n=2$
can be expressed in terms of the transport coefficients that enter  (conformal) viscous hydrodynamics:
\be
\label{eq:p2p2}
{\cal L}_1
=-\frac{2\eta}{t}+\frac{4}{3t^2}(\lambda_1-\eta\tau_\pi)
+ O(1/t^3)
\ee
and
\be
\label{eq:p2p4}
{\cal L}_2=\frac{4}{3t^2}(\lambda_1+\eta\tau_\pi) + O(1/t^3)\,,
\ee
where $\tau_\pi$, and $\lambda_1$ are second order transport coefficients \cite{Baier:2007ix}. 
In obtaining the above results, we have used the following 
expressions of these transport coefficients
in kinetic theory (generalizing Eq.~(\ref{eq:shear_viscosity}) for the shear viscosity)
\begin{subequations}\label{transpcoeff}
\begin{align}
\eta\tau_\pi=&-\frac{T^2}{15}\int_{\t p}\; \t p^5 \t \chi_p \t C_p\,,\\
\lambda_1+\eta\tau_\pi=&\frac{2T^2}{105}\int_{\t p}\; \t p^6 \t \chi_p'\t C_p\,.
\end{align}
\end{subequations}
Again, we emphasize that the momentum dependence of the relaxation time
only affects the values of these transport coefficients, as given by Eqs.~(\ref{transpcoeff}),  but it does not alter the  form of relations such as those given in \Eq{eq:p2p2} and \Eq{eq:p2p4}.

Viscous corrections in hydrodynamics get more 
complicated in higher orders, which makes it more 
involved to derive an explicit correspondance between transport coefficients and the 
 $p^2$-moments associated with higher order Legendre 
polynomials. 
Nevertheless,
it is  possible to generalize the analyses in \Eqs{eq:p2p2}
and (\ref{eq:p2p4}) to higher orders, at least for the 
leading terms. Indeed, it can be shown that, after  linearizing the Boltzmann equation in order to construct 
 the gradient expansion, there is a term with highest power in momentum which appears on 
the left hand side of the Boltzmann equation as a result of
applying $P^\mu \partial_\mu$ to $f_{eq}(P\cdot U/T)$ iteratively.
This highest power  
appears also as the leading contribution to the coefficient of
the Legendre polynomial $P_{2n}(\cos\theta)$ in $\delta f$,
through contraction of irreducible tensors,
\be
\label{eq:tensor_contract}
p^{\mu_1}p^{\mu_2}
\ldots
p^{\mu_{2n}}
\bra \sigma_{\mu_1\mu_2}\ldots\sigma_{\mu_{2n-1}\mu_{2n}}\ket
\propto \frac{p^{2n}}{t^n}P_{2n}(\cos\theta)\,.
\ee
Therefore, one has
\be
\label{eq:cn0}
{\cal L}_n= \frac{1}{t^{n}}\times c_{n}
+O(1/t^{n+1}) \,.
\ee
The coefficient $c_{n}$ is expected to be identified to some 
combinations of
transport coefficients of order $n$, since  
 $1/t^n$ represents an $n$-th order contribution in the 
gradient expansion with a definite angular structure. 
For a conformal and classical 
gas, $c_n$ can be analytically determined when the relaxation time is linear in $p/T$ 
(quadratic ansatz),
\be
\label{eq:cn}
c_n =(-1)^n\frac{(2n)!}{(4n+1)!!} \Gamma(2n+4) \left(\frac{\eta}{s}\right)^n
\frac{T^{4-n}}{2\pi^2}\,.
\ee 
One may regard \Eq{eq:cn} as an analytical prediction of the 
$n$-th order transport
coefficient of a conformal system, with $\eta/s$ a constant input.
In particular, one easily verifies  that $c_1=-2\eta$,  as expected.
More details regarding \Eq{eq:tensor_contract} 
and \Eq{eq:cn} are given in \App{app:tensor_contract}.

\section{Evolution of the ${\cal L}_n$ moments in expanding systems}
\label{sec:sec3}

We now  demonstrate how the ${\cal L}_n$ moments evolve in the pre-equilibrium stage of heavy ion collisions
by calculating them using  kinetic theory. To do so, we solve the
Boltzamnn equation (\ref{eq:kinetic}) for a boost invariant system. 
In line with 
the color-glass  picture  (CGC)~\cite{Gelis:2010nm}, we take an initial 
momentum distribution function of the form
\be
\label{eq:cgc_ic}
f(t_0,p_T,p_z)=f_0 \Theta\left(Q_s-\sqrt{\xi^2 p_z^2+ p_T^2}\right)\,,
\ee
where $f_0$ is a free parameter characterizing the typical 
initial gluon occupation number, and $Q_s$ is the saturation momentum. 
We assume that the 
kinetic  description applies at an initial time $t_0\sim1/Q_s$. In this work, we choose a 
value  $f_0=0.1$, for which the approach to equilibrium occurs smoothly, without encountering Bose-Einstein condensation~\cite{Blaizot:2013lga,Blaizot:2014jna}\footnote{Including the inelastic processes would change the behavior of the distribution function at very small momenta, as discussed in \cite{Arnold:2006fz}, and analyzed in detail recently in \cite{Blaizot:2016iir}. However, because of the weighting factor $p^2$ in their definition, the moments ${\cal L}_n$ are not sensitive to this particular region, and we do not expect that including inelastic, number changing, processes would alter the main conclusions of the present paper}. Especially, with this small initial occupancy, variation of particle number would have negligible effect regarding the $2\leftrightarrow 2$ QCD
matrix elements within a small angle apprximation.
 The parameter 
$\xi$ controls the initial momentum anisotropy.
For  $\xi>1$, one has initially $\P_L/\P_T<1$.
Throughout this work, we take for definiteness a fixed value, $\xi=1.5$, corresponding to an initial
momentum anisotropy $\P_L/\P_T\approx 0.5$. One may check that, initially, the  moments calculated form the momentum distribution (\ref{eq:cgc_ic}) are ordered such that $|{\cal L}_{n+1}|<|{\cal L}_n|$. Furthermore,  all odd order moments
are negative, while even order moments are positive.

We shall present the results of two calculations. 
We start with the  simple case where the collision term is written in the relaxation time approximation.
For a constant value of the relaxation time (linear ansatz), an analytical
solution to the Boltzmann equation can be obtained~\cite{Baym:1984np}. 
For a somewhat more realistic
analysis, 
we then proceed to the numerical solution of the  Boltzmann equations for quarks and gluons,  
considering 2-to-2 scatterings 
among gluons and quarks within a
small scattering angle approximation \cite{Blaizot:2014jna}.

\subsection{Relaxation time approximation}
\label{sec:sec3a}

The Boltzmann equation (\ref{eq:kinetic}),
with the relaxation time approximation for the collision term, reads
\be
\label{eq:relax}
\left[\frac{\partial }{\partial t} 
- \frac{p_z}{t}\frac{\partial}{\partial p_z}\right]f(t, p_T,p_z)
=-\frac{f(t,p_T,p_z)-f_{\rm eq}}{\tR}\,,
\ee
We only require here energy conservation
\footnote{If one would also require conservation of the number of constituents, 
the local equilibrium distribution would also depend 
on a chemical potential \cite{Florkowski:2016qig}. This situation will be considered in the next subsection.}, 
so that the local equilibrium distribution function 
$f_{\rm eq}$ is a Bose-Einstein distribution with a vanishing  chemical potential,
\be
f_{\rm eq}(p_T,p_z)=\frac{1}{\exp(\sqrt{p_T^2+p_z^2}/T)-1}\,.
\ee

\begin{figure}[h]
\begin{center}
\includegraphics[width=0.49\textwidth] {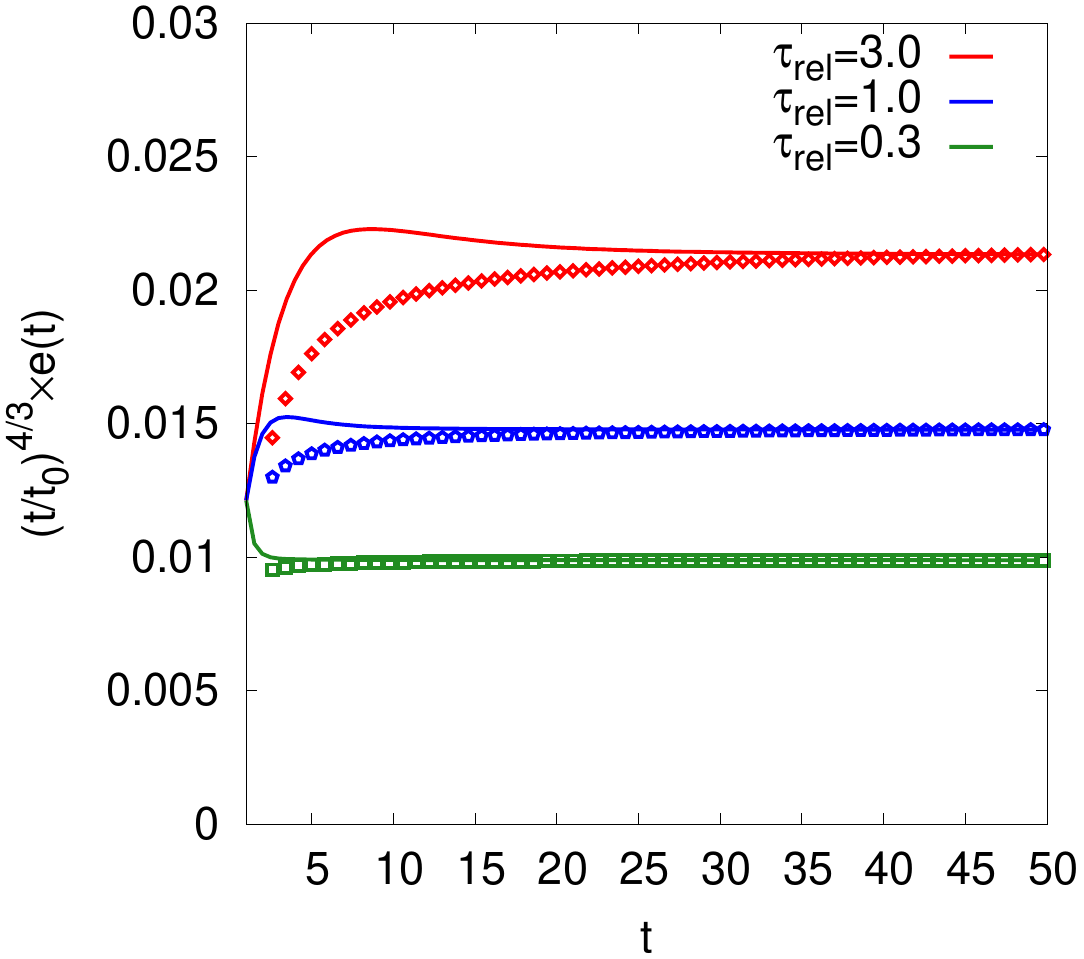}
\includegraphics[width=0.49\textwidth] {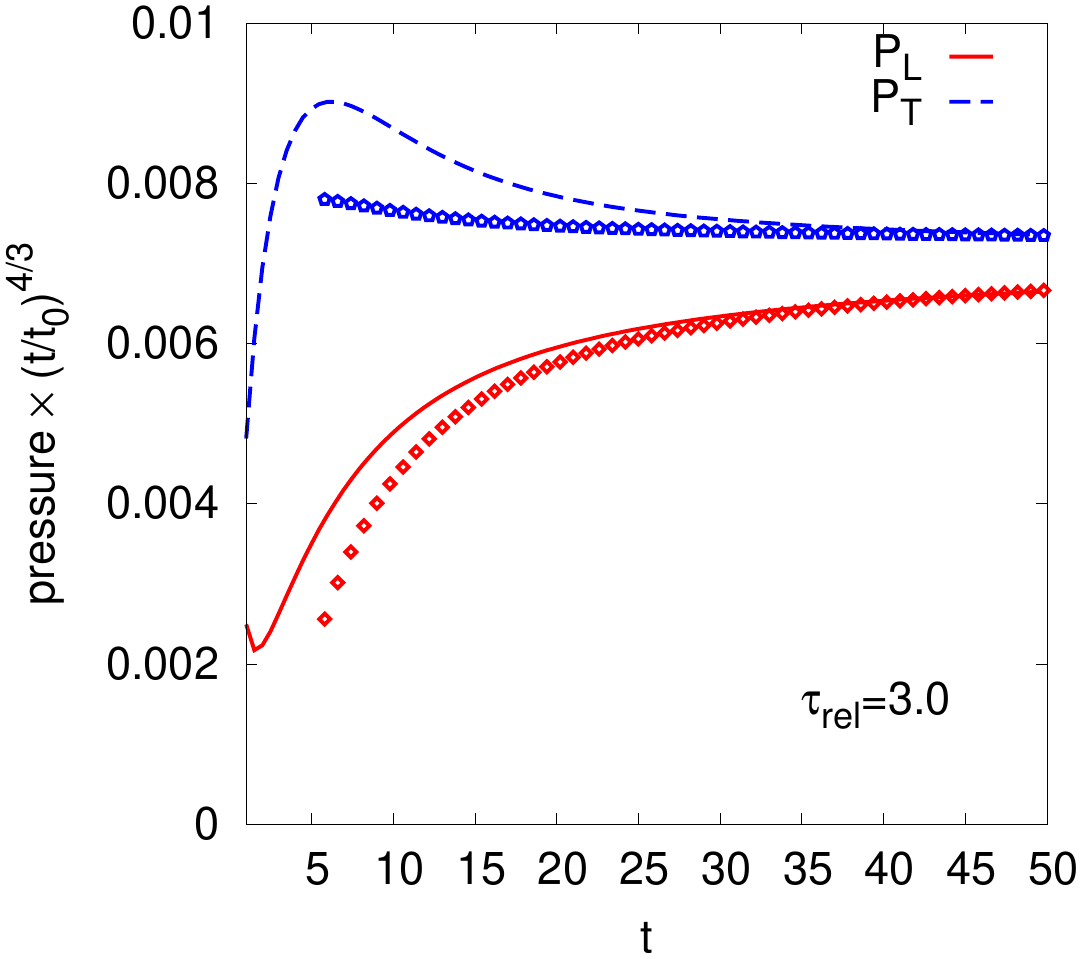}
\caption{ (Color online)
Evolution of the energy density (left), and longitudinal and transverse pressures (right),  obtained by  solving the Boltzmann equation with the relaxation time approximation, \Eq{eq:sol_tR}, full lines. Both quantities are rescaled by a factor $t^{4/3}$ so that they approach constant values in the ideal hydrodynamical regime. The initial difference between the longitudinal and the transverse pressures is due to the parameter $\xi$ in the initial distribution function. The symbols represent the corresponding solutions of the  Navier-Stokes hydrodynamics, started at time $t=50$ (in units of $Q_s^{-1}$) and evolved backwards in time. 
\label{fig:ensrta}
}
\end{center}
\end{figure}

For a constant $\tR$, the solution to \Eq{eq:relax} can be written formally as,
\be
\label{eq:sol_tR}
f(t,p_T, p_z)=
e^{-\frac{t-t_0}{\tR}} f(t_0,p_T,p_z t/t_0)
+\int_{t_0}^t\frac{ dt'}{\tR} e^{-\frac{t-t'}{\tR}} 
\,f_{\rm eq}\left(\sqrt{p_T^2+p_z^2(t/t')^2},t'\right).
\ee
We recognize in the first term in the right hand side  of Eq.~(\ref{eq:sol_tR}) a contribution that represents free-streaming 
from the initial condition $f(t_0,p_T,p_z)$, Eq.~(\ref{eq:cgc_ic}).
 The time dependence of the temperature $T$ in $f_{\rm eq}$ in the second term is fixed by the condition of 
energy conservation 
\be\label{eq:energycons}
e(t)=\int_p p^2f(t,p_T,p_z)=\int_p p^2f_{\rm eq}(t,\p)=\frac{\pi^2}{30} T^4\,.
\ee
This relation, together with Eq.~(\ref{eq:sol_tR}), completely determines the solution. The resulting energy density exhibits the expected transition from the early free streaming regime, where $e(t)\sim 1/t$ (see below),  to the hydrodynamic regime at late times where $e(t)\sim 1/t^{4/3}$.  The evolutions with time of the energy density and the pressures obtained from Eq.~(\ref{eq:sol_tR})  are illustrated in Fig.~\ref{fig:ensrta}, and compared to the solution of first order viscous hydrodynamics (which we also refer to as Navier-Stokes hydrodynamics). As shown by this figure, the energy density is well accounted for by viscous hydrodynamics for times $t\gtrsim 15\tau_{\rm rel}$. This is also the case for the pressure, although in this case, the existence of significant viscous corrections at the latest times is attested by the fact that the longitudinal pressure is not yet equal to the transverse pressure, ${\cal P}_L/{\cal P}_T\simeq 0.9$ at $t=50\,Q_s^{-1}$. 



The moments can be  calculated from the distribution function given in
\Eq{eq:sol_tR}. Consider first  the free-streaming regime ($\tR\rightarrow\infty$). In this case, 
one has
\be
\label{eq:frees_moment}
{\cal L}_n^{\mbox{\tiny FS}}
=\frac{f_0Q_s^4 }{4(2\pi)^2 } 
\left(\frac{t_0}{\xi t}\right)
\mathcal{F}_{n}\left(\frac{t_0}{\xi t}\right),
\ee
where
$\mathcal{F}_n(x)$ is a function defined from the following integral ($0\le x\le 1$)
\be
\mathcal{F}_{n}(x)=\int_{-1}^1 dy \left[1-(1-x^2)y^2\right]^{1/2}
P_{2n}\left(\frac{xy}{\left[1-(1-x^2)y^2\right]^{1/2}}\right)\,.
\ee
This function has the following limits: $\mathcal{F}_n(x)\to \pi P_{2n}(0)/2$ as $x\to 0$, and $\mathcal{F}_{n\ne 0}(x)\to 0$
as $x\to 1$.
Thus, for asymptotically large $t$, $t\gg t_0/\xi$, $\mathcal{F}_n(x)$ 
reduces to a constant, and the moments decay as $1/t$. When $t=t_0/\xi$, 
the moments with $n\ne0$ vanish, 
which implies in particular that they vanish at $t=t_0$ if there is no initial momentum anisotropy  ($\xi=1$). The energy 
density is given by the zeroth moment, with  $\mathcal{F}_0(0)=\pi/2$ and 
$\mathcal{F}_0(1)=2$. To within the slowly varying function $\mathcal{F}_0(t_0/\xi t)$, 
the energy density exhibits the expected behavior in $1/t$. It can also be verified 
that the longitudinal pressure drops rapidly, as $\sim 1/t^2$, so that at times $t\gg t_0/\xi$, 
the distribution function is peaked around $p_z=0$, and  the energy density is 
dominated by transverse degrees of freedom. 
\begin{figure}
\begin{center}
\includegraphics[width=0.55\textwidth] {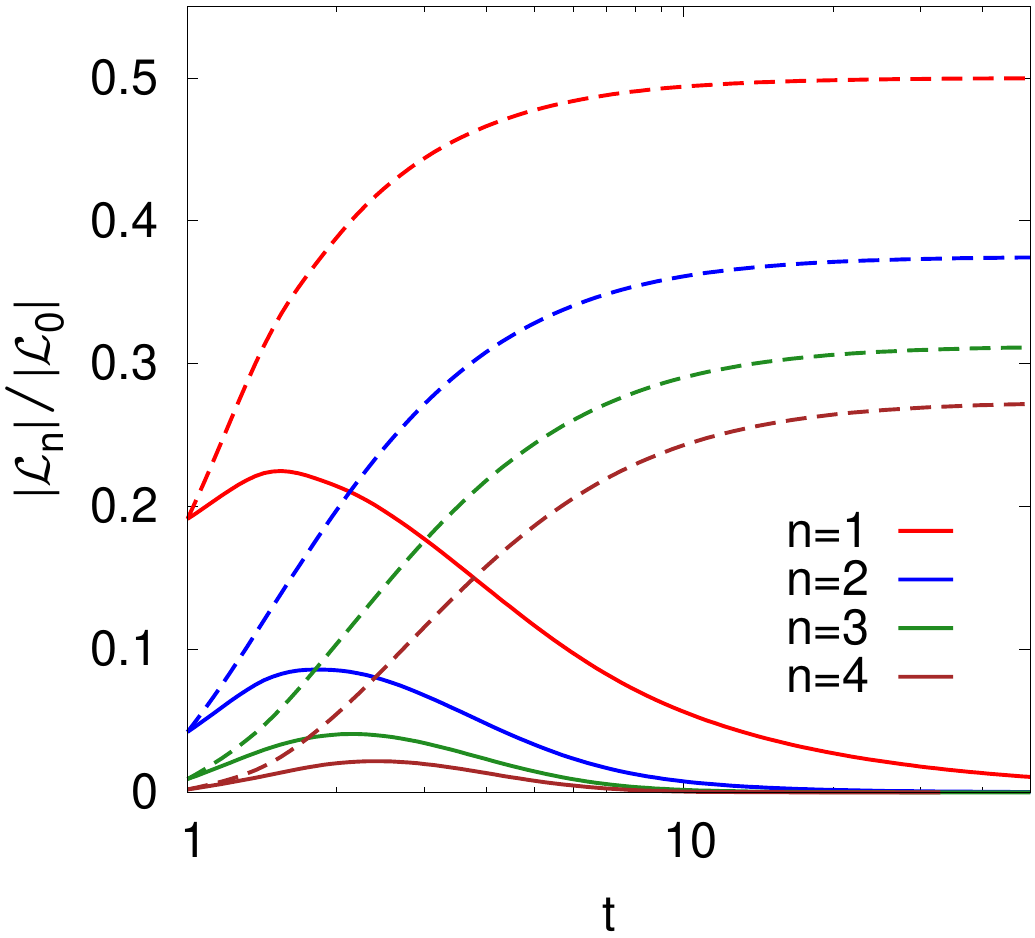}
\caption{(Color online)
Time evolution of the first few moments ($n=1,\cdots, 4$, from top to bottom)  normalized by the energy density, and obtained by 
solving the Boltzmann with the relaxation time approximation. The unit of time is $Q_s^{-1}$, and both the relaxation time $\tR$ and the initial time $t_0$ are set equal to $Q_s^{-1}$, i.e.,   $\tR=Q_s^{-1}=t_0$. 
The dashed lines are the corresponding moments for the free streaming solution, 
\Eq{eq:frees_moment}. We observe that by the time $t\gtrsim 15 \tau_{\rm rel}$, all moments but ${\cal L}_1$ vanish. 
\label{fig:moments_evol_relax}
}
\end{center}
\end{figure}


For a finite $\tR$, \Eq{eq:sol_tR}
leads to
\begin{align}
\label{eq:moment_sol_tR}
{\cal L}_n(t)
=\;e^{-(t-t_0)/\tR}{\cal L}_n^{\mbox{\tiny FS}}
+6\zeta(4)\int_{t_0}^t \frac{dt'}{(2\pi)^2}\frac{e^{-(t-t')/\tR}}{\tR} T(t')^{4}
\left( \frac{t'}{t}\right)
 \mathcal{F}_{n}(t'/t)\,,
\end{align}
where $\zeta(n)$ is the Riemann-zeta function. 
In this equation, the first term represents the contribution of the  free-streaming of
the initial distribution. This is suppressed in a time scale $\tR$, i.e., when collisions start to play a significant role. One thus expects the evolution of the moments to exhibit 
a transition between the free-streaming regime at short time, $t\ll \tR$,  
and the late time regime,  
dominated by collisions and represented by  the second term
in \Eq{eq:moment_sol_tR}. 


\begin{figure}
\begin{center}
\includegraphics[width=0.55\textwidth] {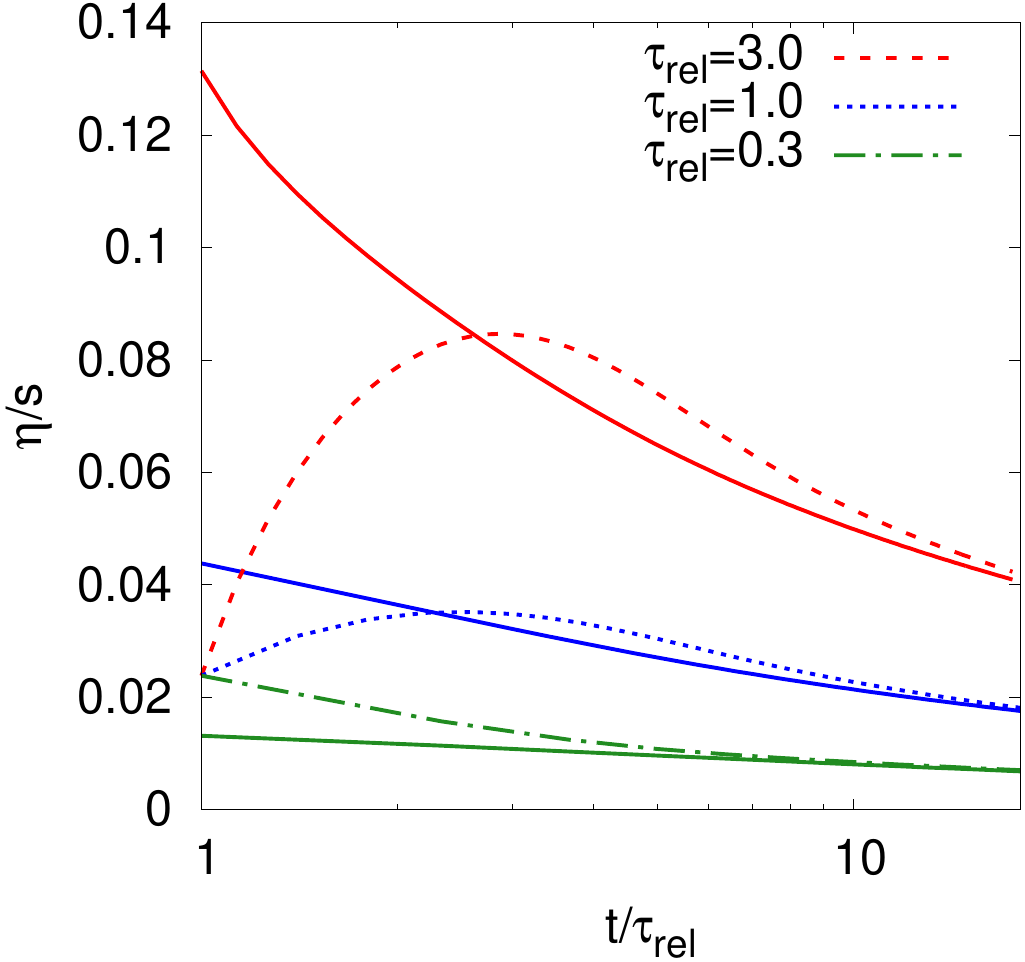}
\caption{(Color online)
Evolution of the   moment ${\cal L}_1$ obtained from the solution of the  Boltzmann equation in the  relaxation time approximation,
plotted in terms of the ratio $-t{\cal L}_1/2s$, with $s $ the entropy density.  According 
to \Eq{eq:p2p2}, this ratio 
can be identified,  in the hydrodynamics regime and in leading order, to $\eta/s$. Three values of the relaxation time are considered $\tR=0.3 ,  1, 3 $ (in units of $Q_s^{-1}$), with the same initial condition (\Eq{eq:cgc_ic}) in the three cases.
The solid lines represent the analytical expression $\eta/s=\frac{1}{5}T\tR$ obtained in the (constant) relaxation time approximation, with the tempearture $T$ determined from Eq.~(\ref{eq:energycons}). Note an unrealistic feature of this approximation which forces $\eta/s$ to vanish at large time.  These curves confirm  that viscous hydrodynamics can provide an accurate description  for $t\gtrsim 15 \,\tau_{\rm rel}$.  
\label{fig:etas}
}
\end{center}
\end{figure}

Figure~\ref{fig:moments_evol_relax} displays the 
evolution of the absolute values of the normalized moments $|{\cal L}_n/{\cal L}_0|$  up to $n=4$ 
(recall  that  ${\cal L}_1$ and ${\cal L}_3$ are negative).
Also shown are the moments of the pure free-streaming
solution, \Eq{eq:frees_moment}, which saturate at late times
to their corresponding asymptotic values determined by $\mathcal{F}_n(0)$,
namely,
\be
\label{eq:ratio_limits}
\frac{|{\cal L}_{2n}^{\mbox{\tiny FS}}  |}
{|{\cal L}^{\mbox{\tiny FS}}_0|}\xrightarrow[]{t\gg t_0}
\frac{(2n-1)!!}{(2n)!!}\,.
\ee
Note that this limit is \emph{independent} of the initial pressure anisotropy paramater $\xi$. 
Clearly, the isotropization of the momentum distribution, signaled by the 
vanishing of the moments, can only be achieved by the collisions. Indeed 
the effect of the collisions starts to be visible around the time scale 
$(t-t_0)\sim \tR$ where they begin to compete with the  free streaming and later 
drive the momentum distribution to isotropy. Another effect of 
the collisions is to reduce the overall magnitude of the higher moments. As 
can be seen on Fig.~\ref{fig:moments_evol_relax}, while the free streaming 
moments evolve towards comparable (within a factor 2) values at late times, 
the hierarchy of moments present in the initial condition is preserved but after 
a time  $t\gtrsim 15\, \tR$ only the moment ${\cal L}_1$ remains significant. 
At this time the evolution of the system is well accounted for by viscous hydrodynamics. \footnote{Note that the hydrodynamical behavior of the moments at large time, that is predicted by Eq.~(\ref{eq:cn0}), is only observed at times later that those considered in Figs.~\ref{fig:moments_evol_relax} and \ref{fig:etas}. This is however of little practical significance, because by that time, these moments have become very small and do not affect much the dynamics.}

This is confirmed by a more detailed study of the  late-time evolution of the moments in \Eq{eq:moment_sol_tR}. 
First we consider the moment ${\cal L}_1$.
According to  \Eq{eq:p2p2},  the ratio between this  moment and the entropy 
density, more precisely 
$-t{\cal L}_1/2s$, approaches $\eta/s$ in
the hydrodynamic regime. This ratio is plotted in \Fig{fig:etas}, where 
we purposely rescaled the time in \Fig{fig:etas} by $1/\tR$ so
that they evolve on the same time scale.  
As time increases, $-t{\cal L}_1/2s$ indeed approaches the 
kinetic theory expectation: $\eta/s=\frac{1}{5}T\tR$, and reaches it for $t\gtrsim 15\,\tR$. 

\begin{figure}[h]
\begin{center}
\includegraphics[width=0.32\textwidth] {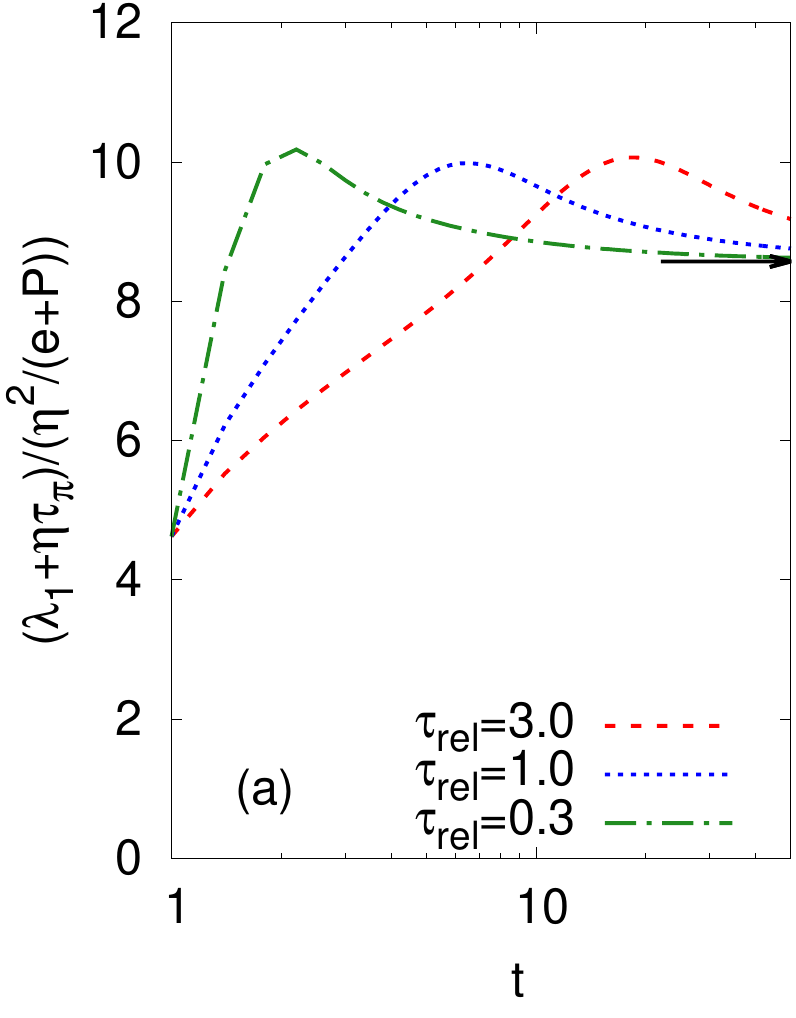}
\includegraphics[width=0.32\textwidth] {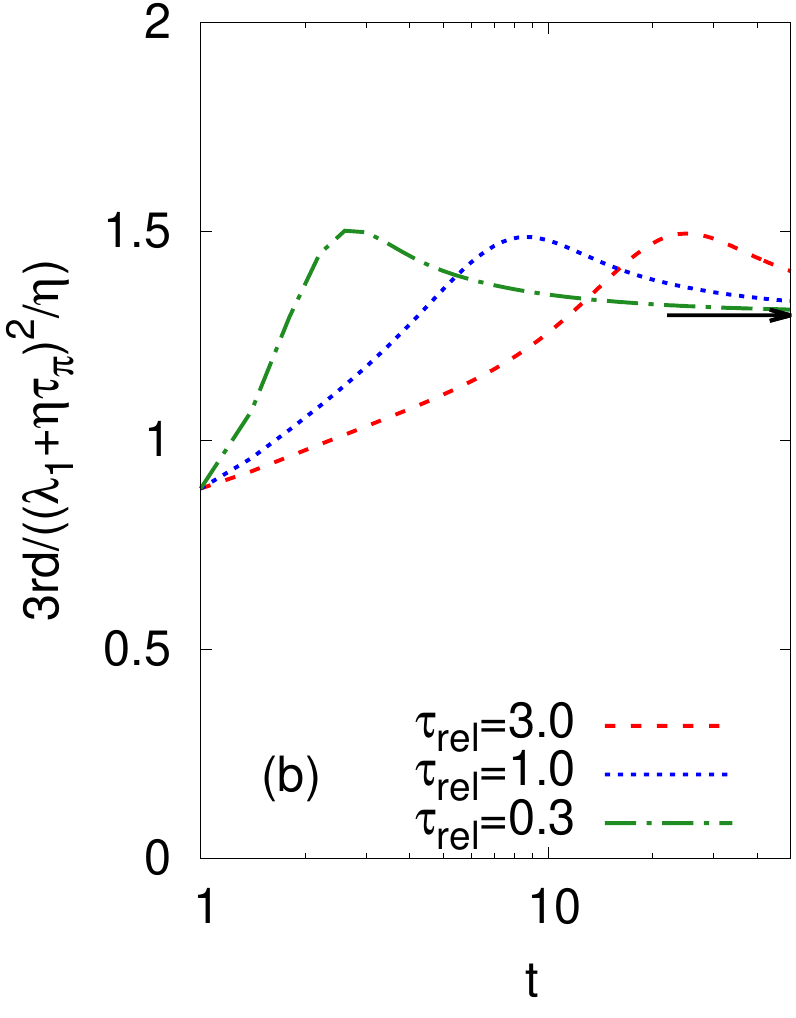}
\includegraphics[width=0.32\textwidth] {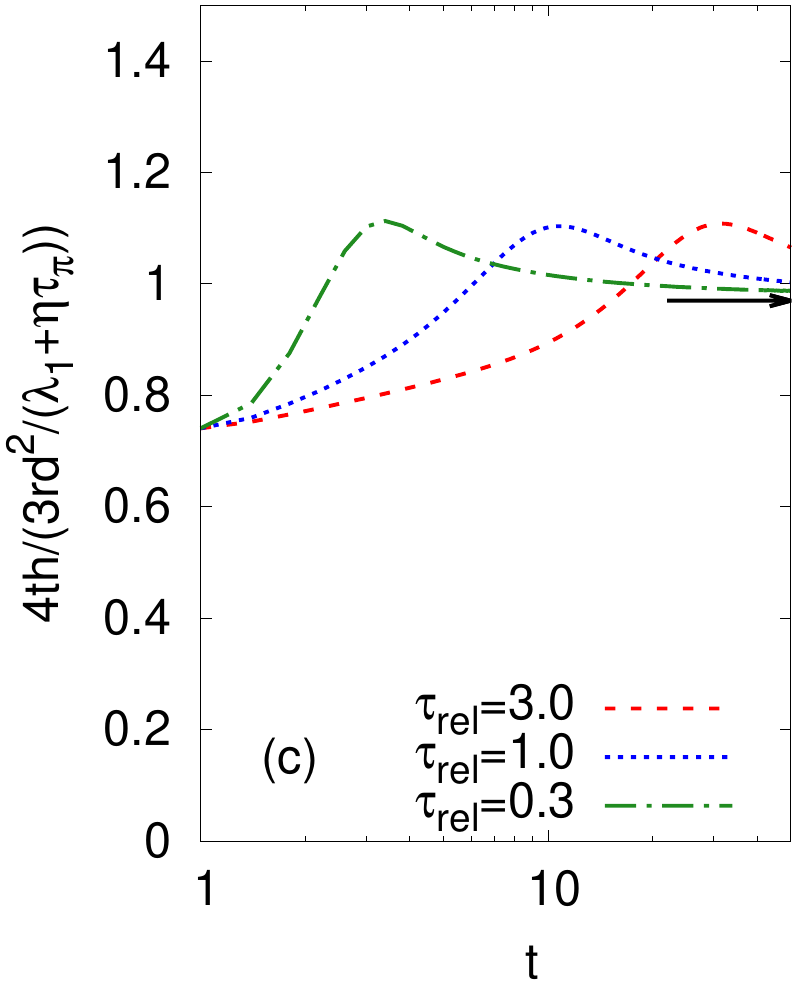}
\caption{ (Color online)
Evolution of higher order moments, plotted in specified combinations which
correspond to dimensionless ratios of transport coefficients in hydrodynamics regime. 
Black arrows indicate asymptotic values. Note that for 
$(\lambda_1+\eta\tau_\pi)/(\eta^2/(e+\P))$, the asymptotic value is consistent with
what was expected from kinetic theory~\cite{York:2008rr}.
\label{fig:higher-orders}
}
\end{center}
\end{figure}

Figure~\ref{fig:higher-orders} completes the discussion and  illustrates the system evolution in terms of higher order moments.
Again, we consider specified dimensionless 
combinations of  moments which are supposed to 
reduce to dimensionless ratios of higher order transport coefficients at late times. 
The following combination
\be
\frac{4{\cal L}_2\times{\cal L}_0}{{\cal L}_1^2}
\rightarrow \frac{\lambda_1+\eta\tau_\pi}{\eta^2/(e+\P)}
\ee 
plotted  in \Fig{fig:higher-orders}(a) 
is related to second order transport coefficients $\lambda_1$ and $\tau_\pi$. In a similar 
way, the combinations
\be
\label{eq:high_trans}
\frac{8}{9}\frac{{\cal L}_1 {\cal L}_3}
{{\cal L}_2^2}
\quad\mbox{and}\quad
\frac{3}{4}\frac{{\cal L}_2{\cal L}_4}
{{\cal L}_3^2}
\ee
that are plotted in \Fig{fig:higher-orders}(b) and (c), respectively,  involve some 3rd order and 4th order
transport coefficients. 
Although these have not yet been 
determined in viscous hydrodynamics\footnote{Note however that the third order coefficients could in principle be extracted from the works on third order viscous hydrodynamics in Refs.~\cite{Jaiswal:2013vta,Grozdanov:2015kqa}. We thank A. Jaiswal for alerting us about this.}
it is nevertheless interesting to estimate their ratios from the asymptotic behaviors: these are the  
black arrows in \Fig{fig:higher-orders}.
Note  that the asymptotic value of 
$(\lambda_1+\eta\tau_\pi)/\eta^2/(e+\P)\approx 8.57$ in \Fig{fig:higher-orders}(a) is
consistent with what one  expects from kinetic theory with a 
linear ansatz for the relaxation time~\cite{York:2008rr}.
The asymptotic values of the ratios in \Fig{fig:higher-orders}(b) and (c) are found to be 
approximately 1.3 and 1.0, respectively. 

\begin{figure}
\begin{center}
\includegraphics[width=0.55\textwidth] {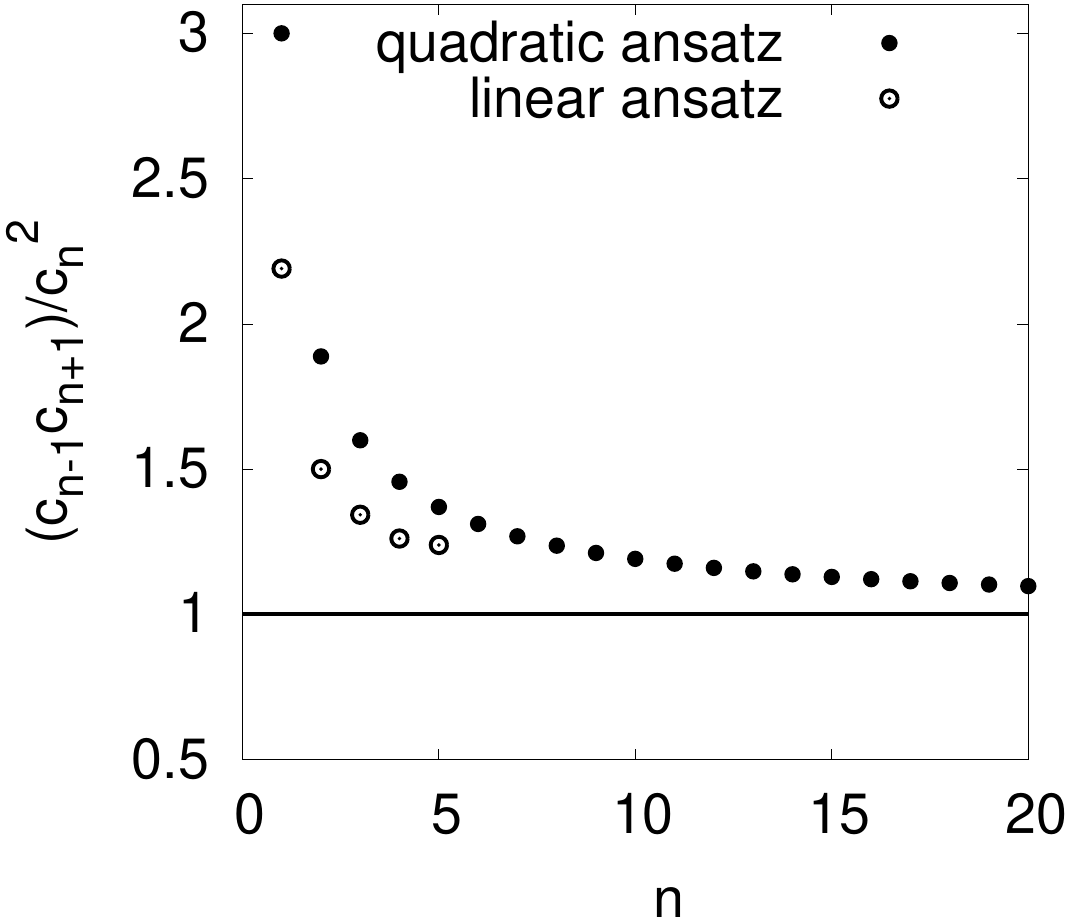}
\caption{Double ratios between transport coefficients, characterized 
in terms from $c_n$'s, for a conformal and classical
gas. The results have been obtained analytically for  the quadratic ansatz of the relaxation time (filled points) and numerically for the 
linear ansatz (open symbols).
\label{fig:cns}
}
\end{center}
\end{figure}
Actually, the combinations plotted in \Fig{fig:higher-orders} are nothing
but double ratios among three consecutive orders of 
transport coefficients (to within simple numerical constants). In terms of the $c_n$'s defined in \Eq{eq:cn0}, these are 
\be
\frac{c_{n-1}c_{n+1}}{c_n^2}=\left(\frac{c_{n+1}}{c_n}\right)\bigg/\left(\frac{c_n}{c_{n-1}}\right)\,.
\ee
Asymptotic values of these double ratios can be obtained analytically in the case of the 
quadratic ansatz for the relaxation time (\Eq{eq:cn}), or numerically for the linear ansatz.  
Results for these two cases are shown in \Fig{fig:cns}
for $n\le 20$ and $n\le 5$, respectively,\footnote{
For the linear ansatz we limit ourselves to  $n\le 5$, since the evaluation of the 
double ratio is challenged by precision in numerical integrations
at asymptotic large $t$ in \Eq{eq:moment_sol_tR}, 
which become less stable for higher orders. 
}
as filled points and open symbols.
For large $n$, the double ratio is seen to 
approach  unity from above, which implies a saturation of transport 
coefficients of asymptotically high orders. This behavior reflects  
the fact  that the gradient expansion leading to viscous hydrodynamics 
is only asymptotic~\cite{Grad:1963,groot1980relativistic} (see also Refs.~\cite{Heller:2016rtz,Denicol:2016bjh} for a recent discussion of this issue within the context of Bjorken expansion).

\subsection{Quark-gluon system with small angle approximation}
\label{sec:sec3b}

We now apply the analysis of the previous section to the evolution of a quark-gluon plasma containing $N_f=N_c=3$ flavors of massless quarks, and introduce separate distributions for quarks ($f_q$) and gluons ($f_g$). We 
consider QCD tree-level 2 to 2 scatterings,  within the  small angle 
approximation.
Following the strategy taken in Ref.~\cite{Blaizot:2014jna}, we reduce 
the coupled Boltzmann equations for the gluon distribution function $f_g$
and the quark distribution $f_q$ to coupled Fokker-Plank equations,
\begin{subequations}
\label{eq:diff_boltz}
\begin{align}
\frac{\partial f_g}{\partial t} + {\bf v}\cdot\nabla_\p f_g&= -\nabla_\p\cdot \mathcal{J}_g + \mathcal{S}_g\,,\\
\frac{\partial f_q}{\partial t} + {\bf v}\cdot\nabla_\p f_q&= -\nabla_\p\cdot \mathcal{J}_q + \mathcal{S}_q\,,
\end{align}
\end{subequations}
where the currents $\mathcal{J}_{q,g}$ and sources $\mathcal{S}_{q,g}$ are given by
\begin{subequations}
\label{eq:current}
\begin{align}
\mathcal{J}_g=&-(4\pi\alpha_s^2\ell)N_c\left[\I_a \nabla_\p f_g + \I_b\frac{\p}{p}f_g(1+f_g)\right]\,,\\
\mathcal{J}_q=&-(4\pi\alpha_s^2\ell)C_F\left[\I_a \nabla_\p f_q + \I_b\frac{\p}{p}f_q(1-f_q)\right]\,,\\
\S_g=&-\frac{N_f}{C_F}\S_q
=\frac{(4\pi\alpha_s^2\ell) C_FN_f \I_c}{p}\left[f_q(1+f_g)-f_g(1-f_q)\right]\,.
\end{align}
\end{subequations}
While the currents are associated to conservation laws (the particle numbers), and hence to elastic processes, the sources arise from the inelastic processes that correspond to quark creations and annihilations. Note that all the  processes considered conserve the total number of particles. 
The quantities $\I_a$, $\I_b$ and $\I_c$ are the following integrals,
\begin{subequations}\label{integrals}
\begin{align}
\I_a&=\int_{p}p^0[N_cf_g(1+f_g)+N_ff_q(1-f_q)],\\
\I_b&=\int_{p}\frac{2p^0}{p}(N_cf_g+N_ff_q),\\
\I_c&=\int_{p}\frac{p^0}{p}(f_g+f_q).
\end{align}
\end{subequations}
For a thermal system in equilibrium, $\I_a/\I_b$ is 
identical to the  temperature. Finally the quantity $\ell$ denotes a  logarithmically divergent integral (the Coulomb logarithm)
\be
\ell=\int_{q_{min}}^{q_{max}} \frac{dq}{q},
\ee
which arises from small angle scatterings. For a QCD 
plasma close to thermal equilibrium, $\ell\approx \ln g^{-1}$ (with $g$ the strong coupling constant). This Coulomb logarithm is treated as a constant in the present work.

 With all the quantities thus
specified, one may check that
the collision kernel in \Eq{eq:diff_boltz} conserves energy-momentum and
constituent number (number of gluons, quarks and anti-quarks), 
which implies that the local equilibrium distribution is, in the 
local rest frame, a Bose-Einstein distribution for the gluons and a
Fermi-Dirac distribution for the quarks and the antiquarks,  with given temperature and chemical potential
(see 
Ref.~\cite{Blaizot:2014jna} for details).

Assuming Bjorken flow, and 
integrating over azimuthal angle, one transforms \Eq{eq:diff_boltz} 
into
\begin{subequations}
\label{eq:diff_boltz1}
\begin{align}
\partial_t f_g-\frac{p^z}{t}\partial_{p^z} f_g=&\frac{1}{p_T}\frac{\partial}{\partial p_T} (p_T J_g^T)
+\frac{\partial}{\partial p^z}J_g^z + S_g\,,\\
\partial_t f_q-\frac{p^z}{t}\partial_{p^z} f_q=& \frac{1}{p_T}\frac{\partial}{\partial p_T} (p_T J_q^T)
+\frac{\partial}{\partial p^z}J_q^z + S_q\,,
\end{align}
\end{subequations}
where $J_g$, $J_q$, $S_g$ and $S_q$ are the currents and sources obtained from Eqs.~(\ref{eq:current}) after a simple rescaling that eliminates the common factor $4\pi \alpha_s^2 {\ell}$. 
That is,  in deriving \Eq{eq:diff_boltz1} we have absorbed the 
 factor $\alpha_s^2\ell$  in a redefinition of the time, 
$t\rightarrow t(\alpha_s^2\ell)$, and we have redefined the integrals (\ref{integrals}) by multiplying them by a factor $4\pi$. With these redefinitions,   the  time scale of the simulation is measured in units of $\tau_s\equiv (\alpha_s^2\ell Q_s)^{-1}$. 
For technical reasons related to the numerical precision of the calculated moments, our simulations
end at $t=400\tau_s$.

We initialize the system at a time $t_0=1/Q_s$, and assume that, at this time, the system contains only gluons, with a distribution function given by \Eq{eq:cgc_ic}. 
In order to be able to compare with the previous subsection, we shall make  
a rough estimate of the relaxation time. To do so, we focus on the gluon sector, and the diffusion  piece of the current (the one proportional to the gradient of the distribution). 
By writing the linearized version of the gluon 
kinetic equation (Eq.~(\ref{eq:diff_boltz})) as  $\delta f_g/\tR \sim \delta f_g/p^2$, we get
\be
\label{eq:tscale1}
\frac{1}{\tR}\sim  4\pi \alpha_s^2\ell N_c^2 \frac{f_0}{p^2} \frac{Q_s^3}{6\pi^2}.
\ee
For $p\approx Q_s$, $N_c=3$,  this yields
\be
\label{eq:trel_est}
\frac{1}{\tR}\sim\frac{1}{\tau_s} \frac{6 f_0}{\pi}\sim 0.2 \quad\mbox{and}\quad  
\tR\sim 5 \tau_s.
\ee

\begin{figure}
\begin{center}
\includegraphics[width=0.55\textwidth] {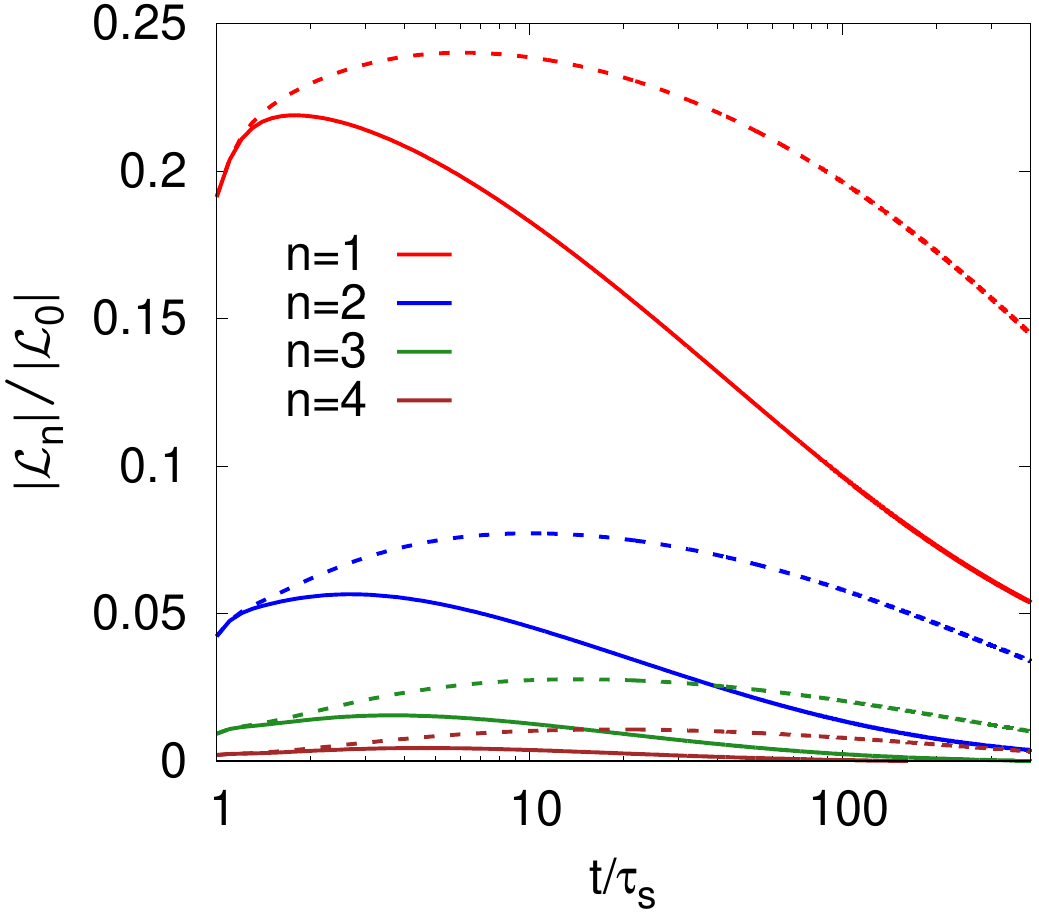}
\caption{ (Color online)
Time evolution of absolute values of the moments ${\cal L}_n$ normalized by the energy density  ${\cal L}_0$,
for a pure gluon system (solid lines) and a quark gluon plasma
 (dashed lines), as obtained from solving the Boltzmann equation
within the small scattering angle approximation. 
\label{fig:moments}
}
\end{center}
\end{figure}
We first plot in \Fig{fig:moments} the absolute values of the 
moments ${\cal L}_n$ divided by  the energy density, for 
$n=1$ up to $n=4$ for a pure gluon system (solid lines) and  a QGP (dashed lines). 
The pattern seen here is analogous to that in Fig.~\ref{fig:moments_evol_relax} 
corresponding to the relaxation time approximation. The curves present a (broad) peak at 
the transition between free-streaming at early times and a collision dominated regime
at late times. Clearly, the 
elastic collisions, treated in the small angle approximation, isotropize the system: all moments tend to vanish
in the collision dominated regime. The higher 
moments are damped on a time scale that decreases with increasing order of the moment. 
Overall, the evolution of moments in \Fig{fig:moments} 
is comparable to that in \Fig{fig:moments_evol_relax}, even semi-quantitatively (to within a factor $\sim 2$) if one takes into account the crude estimate of the relaxation time presented above. 
Note that for a quark-gluon plasma,
the isotropization process is significantly delayed and the transition from the  free-streaming regime  to the
collision dominated regime is slower than for the pure gluon case.\\

\begin{figure}
\begin{center}
\includegraphics[width=0.49\textwidth] {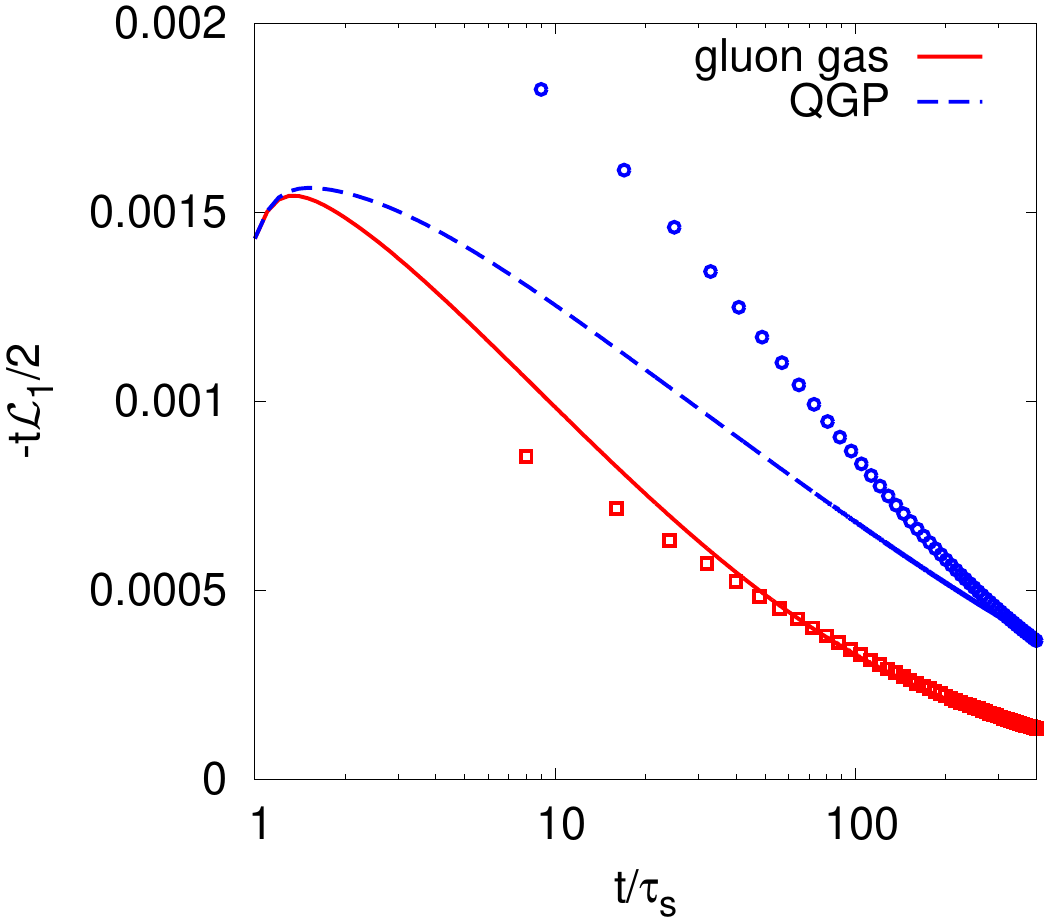}
\caption{ (Color online)
 Evolution of the moment ${\cal L}_1$ obtained by solving the Boltzmann equation for a pure gluon gas (full lines) and a quark-gluon plasma (dashed line). These are compared to the viscous hydrodynamical evolution (symbols), where the initial conditions are fixed at the latest time, $t=400 \,\tau_s$.   
\label{fig:etaB}
}
\end{center}
\end{figure}

In order to compare with hydrodynamics we need to take into account the fact that our Boltzmann equation  conserves the number of constituents. The equation for the energy density needs therefore to be completed by a continuity equation for the number density, which, for Bjorken flow, has the simple solution $n(t) =n(t_0) t_0/t$. This also implies that the local distribution function depends on a temperature and a chemical potential. Details on the hydrodynamical calculations are given in Appendix~\ref{app:hydro}. 
\begin{figure}[h]
\begin{center}
\includegraphics[width=0.49\textwidth] {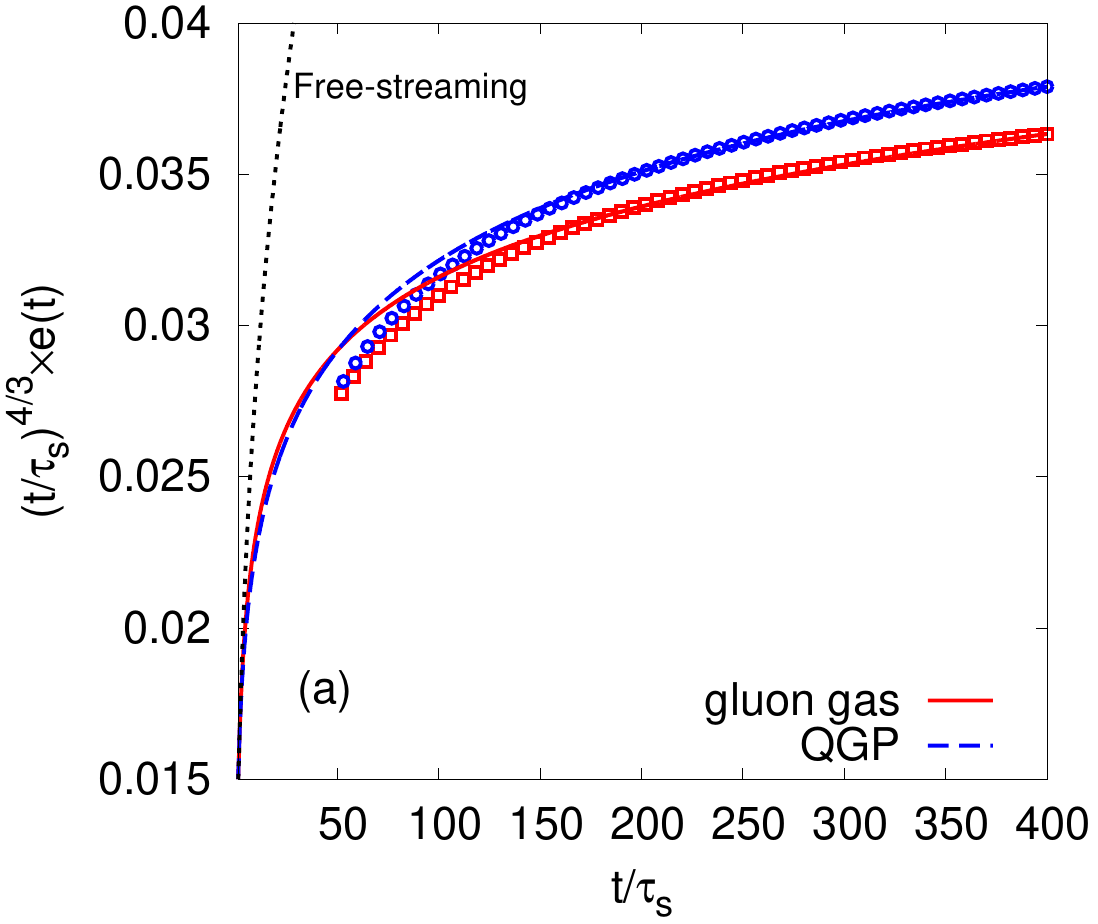}
\includegraphics[width=0.49\textwidth] {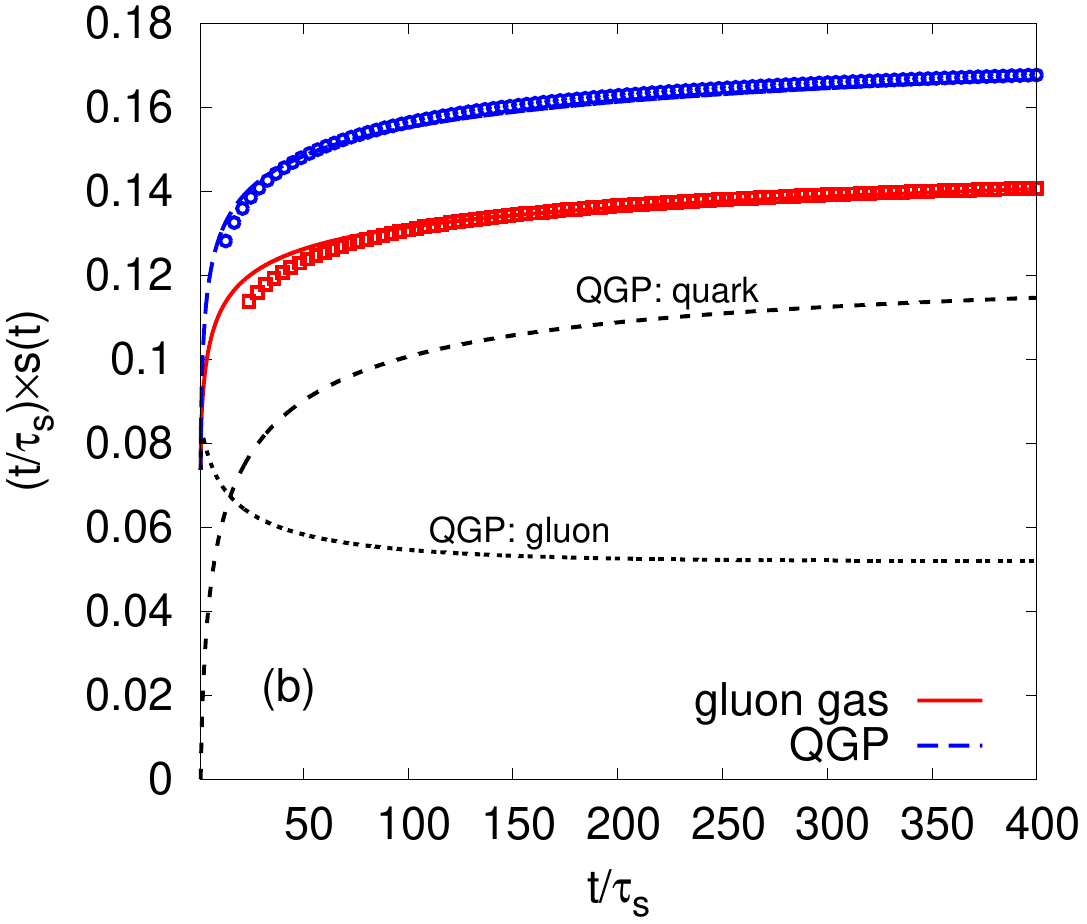}
\caption{ (Color online) Left
(a): Evolution of the energy density (rescaled by $t^{4/3}$) 
of a pure gluon system (red solid line) and 
a QGP (blue dashed line) as obtained by solving the Boltzmann equation within the  small scattering angle approximation, \Eqs{eq:diff_boltz1}. 
Right (b): Evolution of entropy density of a pure gluon gas (red) and
of a QGP (blue). The blue dashed and dotted lines correspond to the gluon and quark
components of a QGP respectively.
The symbols in both (a) and (b) represent the corresponding solutions of  
from Navier-Stokes equations, as described in the text.
\label{fig:ens}
}
\end{center}
\end{figure}

The fact that particle number is conserved can modify the shear viscosity from its value at vanishing chemical potential. When $\mu=0$, 
it is known in weak coupling QCD~\cite{Baym:1990uj,Arnold:2000dr} that 
\be\label{eq:eta_mu0}
\eta(T,0) = \kappa \frac{T^3}{\alpha_s^2 \ell}\,,
\ee 
where $\kappa$ is a constant depending on the number of colors and quark flavors. For instance, for a pure gluon system, $\kappa\approx 0.17$. We do not have (yet) a precise determination of the viscosity appropriate to the present setting. In the present work, we shall therefore rely on 
a crude approximation whereby we use Eq.~(\ref{eq:eta_mu0}) for $\eta$, with $\kappa$ left as an adjustable parameter. 

We initialize the hydrodynamical calculation at $t=400\tau_s$, which corresponds to the largest time for which we can solve accurately the Boltzmann equation. At that time, the 
energy density and the number density are identified with those values obtained from
the solution of the Boltzmann equation.
Since after $t=200\tau_s$,  the 
high order ($n\ge2$) moments are relatively 
less significant, as shown in \Fig{fig:moments}, 
one expects the system evolution to 
follow first order viscous hydrodynamics at   later times, $t\gtrsim 200\tau_{\rm s}$. Note that the ratio $\P_L/\P_T\approx 0.85$ and 
$0.62$ at $t=400\tau_s$ for the pure gluon gas and the QGP,
respectively. This value for the pure glue system (0.85) is to be compared with that obtained in the previous subsection in the relaxation time approximation (0.9), at times where we found the system to be in the (viscous) hydrodynamic regime. 
The evolution of the moment ${\cal L}_1$ is shown in Fig.~\ref{fig:etaB}, and compared to the solution of the  Navier-Stokes hydrodynamics. This comparison allows for the determination of the parameter $\kappa$ in Eq.~(\ref{eq:eta_mu0}). We obtain $\kappa=0.62$ and $0.67$ respectively
 for the pure gluon gas and the QGP. These values are about four times larger than the values expected from weak coupling estimates. These estimates, however,  do not take into account the effect of a non vanishing chemical potential, and the formula (\ref{eq:eta_mu0}) is from that point of view only an approximation in the present context, as already stated. This approximation is sufficient for our purpose: with the viscosity given by Eq.~(\ref{eq:eta_mu0}) the pattern observed in Fig.~\ref{fig:etaB} is similar to that  observed in Fig.~\ref{fig:etas}, 
showing a brief free-streaming 
at early times, followed by a slow  approach to the viscous hydrodynamic regime at late times.



 The evolution
of the energy density obtained from the solutions of \Eqs{eq:diff_boltz}  is presented in \Fig{fig:ens} (a) for 
a pure gluon gas (red solid line) and a QGP (blue dashed line). The energy
density is rescaled by a factor $t^{4/3}$ to be plotted on a visible scale, and also to make clearer the approach to ideal hydrodynamics where the quantity should be constant. One sees that this ideal regime is not reached at $t=400\,\tau_s$, which was also indicated by the non zero value of ${\cal L}_1$ at that time (see Fig.~\ref{fig:etas}).

Figure~\ref{fig:ens} (b) displays  the evolution of the entropy density in the transverse plane. It   goes to a constant  when the ideal hydrodynamical regime is reached. 
As expected in kinetics, a significant amount of entropy is produced 
at early times, when the system starts to re-distribute momenta from the initial 
step function toward a smoother distribution, with in particular a rapid growth of soft modes \cite{Blaizot:2013lga,Blaizot:2014jna}. More entropy is generated in the QGP (blue line), as compared to the pure gluon gas (red line),
due to the additional quark degrees of freedom which, for three flavors, contribute about twice as much as the gluons to the total entropy (chemical equilibrium is nearly reached at $t=400\tau_{\rm s}$). One sees that  Navier-Stokes hydrodynamics  describes well the entropy production already at times  
$t\gtrsim100\tau_s$, for both the gluon gas and the QGP.

\section{Summary and conclusions}
\label{sec:sec4}

In this paper, we have proposed a set of  moments, ${\cal L}_n$, that provide a simple  characterization
of the momentum anisotropy of the momentum distribution function  and take into account 
 Bjorken's symmetry of the  longitudinally expanding quark-gluon plasmas created in relativistic heavy ion collisions. 
At late times, i.e., when the hydrodynamic regime is approached,  these moments can be associated to 
viscous corrections to the energy-momentum tensor, order-by-order in a gradient expansion. The explicit correspondence with the transport coefficients of viscous conformal hydrodynamics has been established. The generalizations to cases
without conformal symmetry can be carried out in a similar way. These moments offer a convenient tool to characterize the onset of hydrodynamics.

The time evolution of the moments was studied first with the help of a Boltzmann equation solved  within  
 a (constant)
relaxation time approximation. This has allowed us to verify the correspondence of the ${\cal L}_n$'s with viscous hydrodynamics  up to order $n=4$.
The onset of hydrodynamics was subsequently analyzed in terms of these ${\cal L}_n$'s
up to $n=4$. The approach to hydrodynamics manifests itself as the decrease of the magnitudes of the ${\cal L}_n$ with $n$ larger than 1.
In fact the moments ${\cal L}_n$ are damped more and more rapidly as $n$ increases, which reflects the fact that  long wavelength modes thermalize faster than short wavelength ones. Note that the preeminence of the moment ${\cal L}_1$ is reminiscent of the arguments used in anisotropic hydrodynamics \cite{Martinez:2010sc,Florkowski:2010cf}.

Similar patterns are observed when we use the more realistic $2\leftrightarrow 2$ QCD matrix elements in the  
Boltzmann equation, which we have solved numerically, within the small scattering angle approximation. 
Again,  
we have determined the evolution of ${\cal L}_n$ 
up to
order $n=4$. 
In the calculations that we have presented, quarks are entirely generated 
via gluon-quark conversion according to tree-level QCD
processes. 
Although quarks are produced on a relatively short time scale, this production is hindered by Pauli blocking effects which tend to delay significantly the thermalization and the onset of hydrodynamics. Finally, the present analysis ignores the inelastic collisions (soft and collinear splitting of gluons) whose role in the isotropization is emphasized in particular in the bottom-up scenario. We defer their study to future work. 

\acknowledgements
We thank Bin Wu and Jean-Yves Ollitrault for helpful discussions. This work has been partially supported by the European Research Council under the Advanced Investigator Grant ERC-AD-267258. LY is supported in part by the Natural Sciences and Engineering Research Council of Canada. JPB also acknowledges the support during the course of this work from the CNRS LIA ``THEP" (Theoretical High Energy Physics) and the INFRE-HEPNET (IndoFrench Network on High Energy Physics) of CEFIPRA/IFCPAR (Indo-French Centre for the Promotion of Advanced Research).

\appendix


\section{Contraction of irreducible tensors and Legendre polynomials}
\label{app:tensor_contract}

The equation (\ref{eq:pps1}) expresses in terms of a Legendre polynomial  the tensorial structure that appears in the leading order gradient expansion of the distribution function. This  can be generalized to higher orders.
In particular, in the $n$-th order viscous corrections,  
the term with the highest power in momentum involves the 
contractions of irreducible tensors of rank $2n$.

Generally, the contraction of a two irreducible tensor of rank $2n$
constructed from the product of $2n$ vectors $A^\mu$ and $2n$ vectors $B_\mu$
can be written in terms of Legendre polynomials
as follows (cf.~\cite{Hess:2015400}),
\be
\label{eq:cont_irre}
{\bra} A^{\mu_1}A^{\mu_2}\ldots A^{\mu_{2n}}\ket
{\bra} B_{\mu_1}B_{\mu_2}\ldots B_{\mu_{2n}}\ket
=|A|^{2n} |B|^{2n} N_{2n} P_{2n}(\hat A\cdot \hat B)\,.
\ee
In this equation, $|A|$ and $|B|$ denote the modulus of the vectors,
while $\hat A$ and $\hat B$ are unit vectors. The constant coefficient is
\be
\label{eq:N2n}
N_{2n}=\frac{(2n)!}{(4n-1)!!}\,.
\ee
As a check, let us apply \Eq{eq:cont_irre} to $P^\mu P^\nu \sigma_{\mu\nu}$. First, 
we note the following identity,
\be
P^\mu P^\nu \sigma_{\mu\nu}\equiv
2 P^\mu P^\nu \bra\nabla_\mu U_{\nu}\ket
=2 \bra P^\mu P^{\nu}\ket \nabla_\mu U_{\nu}
=2 \bra P^\mu P^{\nu}\ket \bra \nabla_\mu U_{\nu}\ket \,.
\ee
We also know that the tensor $\nabla_\mu U_\nu$ with respect to Bjorken flow
has only non-zero component at $\eta\eta$ as 
$\nabla_\eta U_\eta=\tau$,
it is thus able to write equivalently 
\be
\label{eq:eff_tensor_1}
\nabla_\mu U_\nu = R_\mu R_\nu\,,
\ee
with $R_\mu=(0,0,0,\sqrt{\tau})$ as a 4-vector. 
Combining Eqs.~(\ref{eq:cont_irre}) to (\ref{eq:eff_tensor_1})
and making a coordinate transformation from $p^\eta$ to $p^z$,
one gets immediately  \Eq{eq:pps1}
\be
\label{eq:pps2}
P^\mu P^\nu \sigma_{\mu\nu}\equiv
\frac{2 p^2}{\tau}\times \frac{2}{3}\times P_2(\cos\theta)\,.
\ee
The similar procedure can be applied to the contraction of the 
irreducible tensor of rank $2n$ of \Eq{eq:tensor_contract}. We get 
 \begin{align}
 P^{\mu_1}P^{\mu_2}
\ldots
P^{\mu_{2n}}
\bra \sigma_{\mu_1\mu_2}\ldots\sigma_{\mu_{2n-1}\mu_{2n}}\ket
=&2^n\bra  P^{\mu_1}P^{\mu_2}
\ldots
P^{\mu_{2n}}\ket \bra 
\nabla_{\mu_1}U_{\mu_2}\ldots\nabla_{\mu_{2n-1}} U_{\mu_{2n}}
\ket\nonumber\\
=&2^n \bra  P^{\mu_1}P^{\mu_2}
\ldots
P^{\mu_{2n}}\ket \bra 
R_{\mu_1}R_{\mu_2}\ldots R_{\mu_{2n}}
\ket\nonumber\\
=& 2^n \frac{p^{2n}}{\tau^n} N_{2n} P_{2n}(\cos\theta)\,.
 \end{align}

\section{Hydrodynamics with a  with non-zero chemical potential}
\label{app:hydro}

When comparing the solution of the Boltzmann equation  
in \Sect{sec:sec3b} to viscous hydrodynamics, we need to pay attention to the fact that the  QCD 2-to-2 scatterings conserve the number of constituents. As a result the hydrodynamic equations should be complemented with a continuity equation, i.e., we need to solve 
\[
\partial_\mu T^{\mu\nu}=0\,,\qquad
\partial_\mu n^\mu =0\,,
\]
with energy-momentum tensor $T^{\mu\nu}$ given in \Eq{eq:brsss} and 
\be
n^\mu= n U^\mu + I^\mu\,.
\ee
Here, $I^\mu$ represents a viscous correction to the conserved number current. For Bjorken flow, this correction vanishes and, up to first order in
the viscous corrections, the  hydrodynamic equations reduce then to
\begin{subequations}
\label{eq:hydro_ns}
\begin{align}
&\partial_\tau e + \frac{e+\P}{\tau} = \frac{4\eta(T,\mu)}{3\tau^2}\,,\\
&\partial_\tau n + \frac{n}{\tau} = 0\,.
\end{align}
\end{subequations}
These equations are complemented by the equation of state for massless particles, ${\cal P}=e/3\sim T^4$. Furthermore the density and the energy density are related to the temperature and the chemical potential by (for gluons)
\begin{subequations}
\begin{align}
n(T,\mu) =& 2(N_c^2-1) \int_p \frac{p}{\exp(p/T-\mu/T)-1}\,,\\
e(T,\mu) =& 2(N_c^2-1) \int_p \frac{p^2}{\exp(p/T-\mu/T)-1}\,,
\end{align}
\end{subequations}
(with analogous formulae for quarks and antiquarks). Note that the values of the temperature and chemical potential thus determined are not too different from those obtained by a fit of the distribution function calculated from the kinetic equation. 
The second equation (\ref{eq:hydro_ns}) can be solved independently and yields $\tau n(\tau)={\rm cste}$. The equation for the energy density requires the knowledge of $\eta(T,\mu)$. In this work we ignore the explicit dependence of $\eta$ on the chemical potential, and use simply Eq.~(\ref{eq:eta_mu0}) for $\eta$, with $\kappa$ an adjustable parameter. 
The comparison with hydrodynamics is done by fixing ``initial'' conditions at $t=400\tau_s$, and running the equations backwards in time. That is, the initial density and energy density allow us to determine the initial temperature and chemical potential, and hence the viscosity. This process can be repeated to evolve backwards in times steps. The entropy is determined from the   thermodynamic relation $Ts = e+\P-\mu n$, which does not involve explicitly the viscosity. This provides a check of the overall consistency of the calculation.

\bibliography{refsbib}

\end{document}